\documentclass[apj]{emulateapj}
\usepackage{graphicx}
\usepackage{amsmath}
\usepackage{natbib}
\newcommand{\pderiv}[2]{\frac{\partial #1}{\partial #2}}
\newcommand{\pderivL}[2]{\left(\frac{\partial}{\partial #2} + \mathcal{L}\right)#1}
\newcommand{\pderivn}[3]{\frac{\partial^{#3} #1}{\partial #2^{#3}}}
\newcommand{\advec}{\left(\v{u}\cdot\del\right)}
\newcommand{\advecprime}{\left(\v{u}^\prime\cdot\del\right)}
\renewcommand{\v}[1]{{\boldsymbol{#1}}} 

\newcommand{\del}{\v{\nabla}}
\newcommand{\grad}{\del}
\newcommand{\Div}{\del\cdot}
\newcommand{\curl}{\del\times}
\newcommand{\Laplace}{\nabla^2}
\newcommand{\Eq}[1]{Eq. (\ref{#1})}

\newcommand{\Eqss}[2]{Eqs. (\ref{#1})--(\ref{#2})}
\newcommand{\eq}[1]{\Eq{#1}}

\newcommand{\eqss}[2]{\Eqss{#1}{#2}}

\newcommand{\Figure}[1]{Figure~\ref{#1}}
\newcommand{\Fig}[1]{Fig.~\ref{#1}}
\newcommand{\fig}[1]{\Fig{#1}}

\newcommand{\sect}[1]{Sect.~\ref{#1}}

\newcommand{\beq}{\begin{equation}}
\newcommand{\eeq}{\end{equation}}
\newcommand{\beqn}{\begin{eqnarray}}
\newcommand{\eeqn}{\end{eqnarray}}
\newcommand{\xtimes}[2]{{#1}\times{10^{#2}}}
\newcommand{\ttimes}[1]{10^{#1}}
\newcommand{\ksi}{\xi}
\newcommand{\Rey}{\mathrm{Re}}
\newcommand{\urms}{u_{\rm rms}}
\newcommand{\Strain}{\boldsymbol{\mathsf{S}}}

\shorttitle{Fargo MHD}
\shortauthors{Lyra et al.}

\slugcomment{Draft version}

\begin{document}

\title{Orbital advection with magnetohydrodynamics and vector potential}
\author{Wladimir Lyra\altaffilmark{1,2,3,$^\dagger$}, Colin~P.~McNally\altaffilmark{4,3,$^\dagger$}, Tobias Heinemann\altaffilmark{5},  Fr\'ed\'eric Masset\altaffilmark{6}}
\email{wlyra@csun.edu}
\altaffiltext{1}{Department of Physics and Astronomy, California State
  University Northrige, 18111 Nordhoff St, Northridge CA 91130, USA}
\altaffiltext{2}{Jet Propulsion Laboratory, California Institute of Technology, 4800 Oak Grove Drive, Pasadena, CA, 91109, USA}
\altaffiltext{3}{Kavli Institute for Theoretical Physics, Kohn Hall, University of California, Santa Barbara, CA 93106-4030, USA}
\altaffiltext{4}{Astronomy Unit, School of Physics and Astronomy, Queen Mary University of London, Mile End Rd, London E1 4NS, UK}
\altaffiltext{5}{Niels Bohr International Academy, The Niels Bohr Institute, Blegdamsvej 17, DK-2100, Copenhagen \O, Denmark}
\altaffiltext{6}{Instituto de Ciencias F\'isicas, Universidad Nacional Aut\'onoma de M\'exico, Av. Universidad s/n, 62210 Cuernavaca, Mor., Mexico}
\altaffiltext{$^\dagger$}{Both authors contributed equally to this work}

\begin{abstract}

Orbital advection is a significant bottleneck in disk
simulations, and a particularly tricky one when used 
in connection with magnetohydrodynamics. 
We have developed an orbital advection algorithm suitable for
the induction equation with magnetic potential. 
 The electromotive force is split into advection and shear
terms, and we find that we do not need an advective gauge since solving the
orbital advection implicitly precludes the shear term from canceling
the advection term. 
We prove and demonstrate the third order in time accuracy of the scheme.
The algorithm is also suited to non-magnetic problems. Benchmarked
results of (hydrodynamical) planet-disk interaction and the of
the magnetorotational instability are reproduced. 
We include detailed descriptions of the construction and selection of stabilizing dissipations (or high frequency filters) needed to generate practical results.
The scheme is self-consistent, accurate, and elegant in its simplicity, making it particularly efficient for
straightforward finite-difference methods. As a result of the work, the algorithm is
incorporated in the public version of the {\sc Pencil Code}, where it can be
used by the community. 

\end{abstract}

\section{Introduction}
\label{sect:introduction}
In numerical models of gaseous disks in astrophysical systems,
a common challenge is mitigating the difficulties caused by the superposition of the dynamics of interest on a fast azimuthal flow.
In particular,  this applies in accretion disks such as protoplanetary disks, cataclysmic variables, and X-ray binaries.
In disks in hydrostatic equilibrium, the Mach number of the azimuthal flow with respect to the sound speed
is on the order of $h^{-1}$, where $h\equiv H/R$ is the disk aspect ratio,
with $H$ the scale height and $R$ the distance to the central
object. As these disks are thin, the Mach number is usually substantial. 

Additionally, if the disk is magnetized but dominated by thermal pressure,  
then other relevant speeds of information propagation such as the
Alfv\'en and fast magnetosonic speeds will be on the order of
the sound speed or smaller.
Thus, the orbital advection will not just dominate the advection, but
also dominate over the propagation of signals in the flow. 
As a result, special techniques to deal with the orbital advection in such disks have proved very useful for their simulation,
both for ameliorating the negative effects of numerical diffusion during the fast azimuthal advection, and as a way of removing 
the time step constraints imposed on time-explicit integration by the Courant-Fredrichs-Lewy (CFL) condition.

The {\sc FARGO}  code 
\citep[Fast Advection in Rotating Gaseous Objects; ][]{2000A&AS..141..165M} 
introduced a split azimuthal advection, giving rise to the 
popular use of the term ``fargo'' to denote this feature in other schemes.
However, split azimuthal advection has been implemented in a number of
different ways, which can nonetheless be broadly summed up in three categories.
First, there are schemes which directly solve the full equations in terms of the total velocity,
 including the azimuthal advection velocity of the gas on a fixed mesh.
{\sc FARGO}  \citep{2000A&AS..141..165M} is one of these schemes. 
{\sc LA-COMPASS} uses a dimensionally split Lagrangian remap piecewise parabolic method, 
with the advection step arranged as a separate shift by a integer number of cells in the azimuthal sweep \citep{LILI2012}
and allows refined meshes to move across coarse parent meshes.

Second, there are schemes which separate the velocity field into two parts, and dynamically solve only for perturbations to a background flow.
This is the most common in local shearing-box models as opposed to full cylindrical disks, 
including the implementation where the term orbital advection appears to first originate \citep{2008ApJS..177..373J}.
The method implemented for shearing box simulations in the {\sc Pencil Code} \citep{2002CoPhC.147..471B,2010ascl.soft10060B}
is the shearing box orbital advection scheme
Shear Advection by Fourier Interpolation (SAFI), introduced by \citet{2009ApJ...697.1269J}. 
The method of \citet{2001ApJ...551..874L} uses a hybrid approach, enabled by virtue of the directional splitting of the equations. 
In the angular sweep the equation solved is for a fluctuation from the background azimuthal flow, and this sweep is 
sub-cycled in time so that the global time step is only determined by the radial CFL condition.
In  {\sc ATHENA} \citep{2010ApJS..189..142S} the orbital advection implemented is of the type 
which solves for fluctuation velocities and is the first introduction of the 
use of a constrained transport algorithm to solve for the magnetic field in the azimuthal interpolation step.
For {\sc PLUTO} \citep{2010ascl.soft10045M} the formulation introduced in \citet{2012A&A...545A.152M} gives 
a total angular momentum conserving 
formulation of the procedure, solving for velocity fluctuations on a cylindrical grid in a Godunov-type method.

In the third category we place methods which use a moving or variable mesh, or no mesh at all, to 
achieve Lagrangian or partially Lagrangian properties of the method.
{\sc DISCO} \citep{2016ApJS..226....2D,2016ascl.soft05011D} uses a Godunov method implemented on a moving mesh, 
arranged as a series of concentric sliding rings which follow the orbital motion of the disk. 
A novel constrained transport formulation 
adapted to this moving mesh is used for magnetohydrodynamics (MHD).
Alternatively, more general  approaches can be used, either with moving
Voronoi meshes \citep{2010MNRAS.401..791S,2014MNRAS.445.3475M} 
or meshless methods \citep[e.g.,][]{2012ApJS..200....6M,2012ApJS..200....7M} 
which, by allowing the discretization elements to follow the flow in general, also allows them to follow the orbital motion.
Additionally, one might achieve a partly Lagrangian method by allowing the mesh to shear with the background flow, 
and remap to the original coordinates not within each time step, 
but after a significant number and before the mesh becomes too distorted.
This approach is used in some pseudospectral codes \citep{2004A&A...427..855U,2005A&A...444...25L}.

In all methods, the treatment of the centrifugal, Coriolis, and gravitational 
source terms in the radial direction can be critical.
In a flow in steady-state, the orbital motion of the disk results in the partial cancellation 
of the radial component of the gravitational force and (depending on the frame and the full or fluctuation velocity)
the centrifugal and Coriolis force terms; the remainder is balanced against the radial pressure gradient.
In many studies the ability to hold the physical equilibrium state numerically steady for many dynamical times is important.
Thus many methods with lower spatial accuracy need to either analytically cancel radial force terms, or at least 
ensure the mentioned forces are treated in a similar enough  way (e.g., on the same substep of an 
operator split method) that they cancel with sufficient accuracy.

It is worth emphasizing that, across the range of methods
mentioned above, the problem of advancing the 
magnetic field with the disk motion has taken three distinct approaches.
Solving directly for the magnetic field in a ZEUS-family operator-split code, 
\citet{2008ApJS..177..373J} constructed a divergence-free slope-limited remap procedure 
for the azimuthal transport step. This was greatly simplified in the context of a Godunov method in \citet{2010ApJS..189..142S} 
 by applying a constrained transport method to the azimuthal transport of the magnetic field.
{\sc DISCO} \citep{2016ascl.soft05011D} uses a novel constrained transport formulation 
adapted to its mesh of sliding rings.
Using the magnetic vector potential in the {\sc Pencil Code}, \citet{2009ApJ...697.1269J} took
advantage of the Keplerian gauge mandatory for shearing boxes so that 
the vector potential field could be simply shifted and interpolated along the grid.

In this paper, we introduce a novel method that we implemented for the
purpose of this work in the {\sc Pencil Code}. {\sc Pencil} is a high-order finite-difference 
method code using point collocated values, here on a cylindrical mesh,
with a method of lines approach using sixth-order finite-difference discretization in space,
to which a third-order time integration scheme is then applied.
Our method is classified in the first category above, solving for the entire flow, but splitting the 
orbital advection operator, discretizing it in space with a spectral method, and integrating
 it in time along with the rest of the operators in the original Runge-Kutta time integration.
This smoothly inherits the accuracy properties of the base scheme. 
We introduce an appropriate splitting of the vector potential evolution equation so 
that this field needs only to be shifted and interpolated across the grid to enable MHD.
We also extensively 
discuss some choices for the stabilizing dissipation operators which must be included in such a 
scheme for stability, as the use of a cylindrical mesh means significant attention must be paid to this 
matter. Our method is closely related to that of  
\citet{2009ApJ...697.1269J} (SAFI) and, importantly, our proof of third-order accuracy in time applies to it too.
However, as our method is developed for cylindrical-global models, 
the model equations solved are different, using the entire velocity, not simply fluctuations in a shearing box,
and the choice of stabilizing dissipation must be different due to the cylindrical-polar grid.

This paper is organized as follows. In \sect{sect:model-equations} we
describe the model equations. In \sect{sect:thirdorderproof} we
analytically prove the scheme's 3rd-order accuracy in time. Tests are
given in \sect{sect:tests} where we also describe and test in detail 
the dissipation operators needed to stabilize the scheme for these practical problems, 
finally leading to our summary in
\sect{sect:conclusions}.

\section{Model equations} 
\label{sect:model-equations}

We first consider the MHD inviscid equations

\begin{eqnarray}
  \pderiv{\rho}{t}  &=& -\advec\rho -\rho{\Div\v{u}}, \label{eq:continuity}\\
  \pderiv{\v{u}}{t} &=& -\advec\v{u} -\frac{1}{\rho}\del{p} - \del\varPhi + \frac{\v{J}\times\v{B}}{\rho}, \label{eq:navier-stokes}\\
  \pderiv{s}{t} &=& -\advec s + \frac{1}{\rho T} \left(\mathcal{H} - \mathcal{Q}\right), \label{eq:entropy}\\
  \pderiv{\v{B}}{t} &=& \curl{ \left(\v{u}\times\v{B} - \eta\mu_0\v{J}\right)} \label{eq:induction}\\
  p&=&\rho c_s^2/\gamma\label{eq:eos}.
\end{eqnarray}

\noindent where $\v{u}$ is the velocity, $\v{B}$ is the 
magnetic field, $\v{J}=\mu_0^{-1}\curl{\v{B}}$ is the current density,
and $\mu_0$ is the magnetic constant; $p$ is the 
pressure,  $c_s$ is the sound speed, and $\gamma$ is 
the adiabatic index. The gravitational potential $\varPhi=-GM_\star/r$ where 
$G$ is the gravitational constant, and $M_\star$ is the stellar
mass. We restrict ourselves to the cylindrical approximation where the
vertical stratification is ignored, so $r$ is the cylindrical radius. \Eq{eq:entropy} is the energy equation, where $s$ stands for entropy, $T$ is the temperature, $\mathcal{H}$ is a heating term, and $\mathcal{Q}$ is a cooling term. 

Instead of solving for the magnetic
field, in the {\sc Pencil Code} we write the induction equation in terms of the magnetic
potential $\v{A}$, where $\v{B}=\curl\v{A}$. Removing the curl on both
sides, the induction equation for $\v{A}$ is 

\beq
 \pderiv{\v{A}}{t} = \v{u}\times\v{B} -\eta\mu_0 \v{J} 
 \label{eq:A3-induction}
\eeq

This formulation is not adequate for use with an orbital advection
algorithm since it lacks an explicit advection term. Instead, we
expand the electromotive force 

\beqn 
\v{u}\times\v{B} &=& \ksi_{ijk} u_j B_k \nonumber \\
&=& \ksi_{ijk}\ksi_{kpm} u_j \partial_p A_m \nonumber  \\
&=& (\delta_{ip}\delta_{jm} -\delta_{im}\delta_{jp} ) u_j \partial_p A_m \nonumber  \\
&=& - u_j (\partial_j A_i - \partial_i A_j)  \nonumber \\
&=& -\left(\v{u}\cdot\del\right) \v{A}  + \v{u} \cdot \left(\grad \v{A}\right)^T 
\label{eq:expand_uxb}
\eeqn

\noindent to find the advective term $u_j \partial_j A_i$. We use this
formulation to apply the orbital advection algorithm to the magnetic potential. 

The remainder of the algorithm is as the original FARGO, with the care
of performing it as part of a third-order Runge-Kutta timestepping, that
has three stages. 

At the beginning of the first stage the average azimuthal
velocity is defined as a 2D array (in the radial and vertical
coordinates), $\bar{u}_\phi(r,z)$. 

At every stage the residual velocity is computed 

\beq
\v{u}^\prime = \v{u} -\bar{u}_\phi (r,z)\v{\hat{\phi}}
\eeq

\noindent and that becomes the advection azimuthal velocity for the Courant
condition. The modified equations of motion are 

\beqn
  \pderivL{\rho}{t}  &=& -\advecprime\rho -\rho{\Div\v{u}}, \label{eq:continuity2}\\
  \pderivL{\v{u}}{t} &=& -\advecprime\v{u} -\frac{1}{\rho}\del{p} - \del\varPhi + \frac{\v{J}\times\v{B}}{\rho}, \label{eq:navier-stokes-rad}\\
  \pderivL{s}{t} &=& -\advecprime s + \frac{1}{\rho T} \left(\mathcal{H} - \mathcal{Q}\right), \label{eq:entropy2}\\
\pderivL{\v{A}}{t} &=&   -\advecprime\v{A} +  \v{u} \cdot \left(\grad\v{A}\right)^T - \eta\mu_0 \v{J} \label{eq:induction2}
\end{eqnarray}

with 

\beq
\mathcal{L} = \frac{\bar{u}_\phi (r,z)}{r}\frac{\partial}{\partial\phi}
\eeq

This $\mathcal{L}$ term is interpolated in Fourier space as the
SAFI algorithm does with the Keplerian advection in the shearing
box \citep{2009ApJ...697.1269J}.

\subsection{Inertial Terms}

The $\mathcal{L}$ operator acting on a vector $\v{\psi}$ produces inertia terms of the form
$r^{-1}\bar{u}_\phi \psi_i \v{\hat{e}}_j$. These combine with similar 
inertial terms $r^{-1}u_\phi^\prime \psi_i \v{\hat{e}}_j$ produced by
$\advecprime\v{\psi}$ so that the inertia terms on $\v{u}$ and $\v{A}$
are unaltered with respect to the implementation without orbital
advection, i.e., they carry the full velocity.  

The term $\v{u} \cdot \left(\grad\v{A}\right)^T$ produces the inertia
terms $r^{-1}(A_r u_\phi - A_\phi u_r) \v{\hat\phi}$. 


\subsection{The pseudo-advective gauge} 

\begin{figure}
\includegraphics[width=\columnwidth]{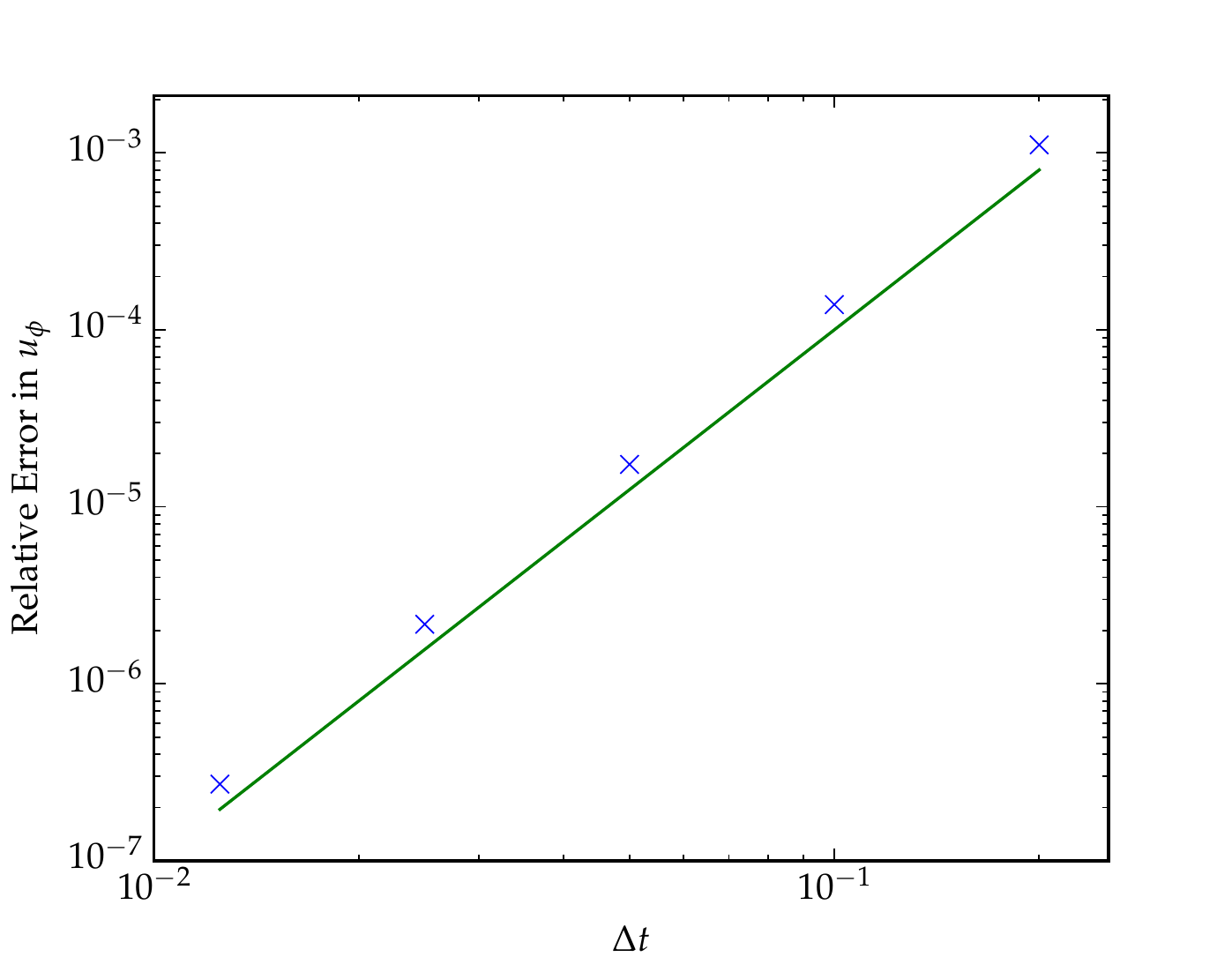}
\caption{Demonstration of the third-order accuracy of the time integration algorithm.
Relative errors of the solution for each time step size are drawn as blue crosses, 
and a $\Delta t^3$ reference slope is shown demonstrating that the numerical solution converges at third order with respect to the time discretization.}
\label{fig:timetest}
\end{figure}

In \eq{eq:expand_uxb}, the term $u_j \partial_i A_j$, if
implemented in connection with the full advection term, would
contribute terms that simply cancel the advection term numerically. 
The reason why \eq{eq:expand_uxb} works with orbital advection is that we split the velocity in the advection
term 

\beq
\partial_t A_i = - \bar{u}_j \partial_j A_i - u^\prime_j \partial_j
A_i + u_j\partial_i A_j \\
\label{eq:lyra-mncally-pseudogauge}
\eeq

\noindent and implement the first term in the right-hand side implicitly. Other
formulations \citep{Brandenburg+95,2010MNRAS.401..347B,2011PhPl...18a2903C},  
without orbital advection, explicitly summon an advective
gauge, transforming the  $u_j \partial_i A_j$ term according to 

\beq
u_j \partial_i A_j =  \grad{(\v{u}\cdot\v{A})} - A_j \partial_i u_j
\eeq

\noindent and because the curl of a gradient is zero, the gradient term has no
bearing in the magnetic field. It can be removed by a gauge
transformation, allowing one to write 

\beq
 \partial_t A_i = -u_j \partial_j A_i  -  A_j \partial_i u_j 
 \label{eq:advective-gauge}\\
\eeq

\noindent which is the advective gauge. Notice that, essentially, the
advective gauge exchanges gradients of $\v{A}$ for gradients of
$\v{u}$. Unfortunately, although analytically elegant, this formulation
is not numerically stable. This is because, although the gradient term is
analytically irrotational, in practice the numerical cancelation is not 
perfect. The unphysical residual of the gradient term, although negligibly small at first, grows 
unboundedly in time, to the point that it eventually 
dominates over the (physical) solenoidal component, leading to a spurious accumulation of power in
small scales of the magnetic potential \citep{2011PhPl...18a2903C}. These in turn lead to inaccurate magnetic fields and 
numerical instabilities at the grid scale. For this reason we work
with \eq{eq:lyra-mncally-pseudogauge},
instead of the advective gauge. This choice does
not have the properties of an advective gauge, but it has proved
well suited to the orbital advection algorithm. For this reason we
call the evolution equation of the vector potential in
\eq{eq:induction2} and \eq{eq:lyra-mncally-pseudogauge} a pseudo-advective gauge.

\section{Third-order accuracy in Time}
\label{sect:thirdorderproof}

In this section we prove analytically that the scheme of
\eqss{eq:continuity2}{eq:induction2} retains the third-order accuracy of
the original formulation without orbital advection.  We consider a partial differential equation of the form 
\beq
\left(\pderiv{}{t} +\mathcal{L}\right) u = f 
\label{eq:orig}
\eeq
\noindent where $u(x,y,t)$ and $f(x,y,t)$ are functions of space and time and
$\mathcal{L}$ is a linear differential operator that we assume not to
depend explicitly on time.

The solution to the homogeneous part of \eq{eq:orig} may be written as 
\beq
u(x,y,t) = \mathcal{Q}(t) u(x,y,0)
\eeq
where 
\beq
\mathcal{Q}(t) = {\rm exp}{\left(-t \mathcal{L}\right)}
\label{eq:Q}
\eeq
is a linear operator. We now introduce new dependent variables $\tilde{u}$ and $\tilde{f}$
through 
\beqn
u(x,y,t) &=&\mathcal{Q}(t) \tilde{u}(x,y,t) \nonumber\\
f(x,y,t) &=& \mathcal{Q}(t) \tilde{f}(x,y,t)
\label{eq:trans}
\eeqn

Plugging \eq{eq:trans} in \eq{eq:orig} we have 
\beq
\pderiv{}{t}\left[\mathcal{Q}(t)\tilde{u}\right] +\mathcal{L}\mathcal{Q}(t) \tilde{u} =
\mathcal{Q}(t) \tilde{f} 
\label{eq:origtrans1}
\eeq
and given \eq{eq:Q} we have 
\beq
\left[\mathcal{Q}(t)\pderiv{\tilde{u}}{t} - \mathcal{L}\mathcal{Q}(t)\tilde{u}\right] +\mathcal{L}\mathcal{Q}(t)\tilde{u} =
\mathcal{Q}(t) \tilde{f} 
\label{eq:origtrans}
\eeq
which eliminates $\mathcal{L}$, leaving only 
\beq
\pderiv{\tilde{u}}{t} = \tilde{f}
\label{eq:transformed}
\eeq
 The effect of the transformation \eq{eq:trans} is to formally
 eliminate the linear operator term in \eq{eq:orig}. With $\mathcal{Q}(t)$ being the advection by $\bar{u}_\phi(r,z)$,
this translation may conveniently be carried out in Fourier space as
in the SAFI algorithm. 

\subsection{The RK3-2N Time Integration Scheme}

The {\sc Pencil Code} uses a time-stepping scheme that, when applied to \eq{eq:transformed},
reads 

\beqn
\tilde{\omega_i} &=& \alpha_i \tilde{\omega}_{i-1} + \tilde{f}_i\\
\tilde{u}_{i+1} &=& \tilde{u}_i + \beta_i \tilde{\omega}_i\delta{t}
\eeqn

\noindent where $i$ runs from 0 to 2 and $u_{3}  = u(t+\delta t)$. 
The parameters $\alpha_i$ and $\beta_i$ are constant coefficients that 
depend on the integration scheme. We use the values of \cite{Williamson80}, 
$\alpha$=[0,-5/9,-153/128] and $\beta$=[1/3,15/16,8/15].

Given \eq{eq:trans} and also applying the operator $\mathcal{Q}(t)$ to $\omega$

\beq
\omega(x,y,t) = \mathcal{Q}(t) \tilde{\omega}(x,y,t)
\eeq

\noindent we have 

\beqn
\mathcal{Q}(-t_i){\omega_i} &=& \alpha_i \mathcal{Q}(-t_{i-1}){\omega}_{i-1} + \mathcal{Q}(-t_i){f}_i\\
\mathcal{Q}(-t_{i+1}){u}_{i+1} &=& \mathcal{Q}(-t_i){u}_i + \beta_i \mathcal{Q}(-t_i){\omega}_i\delta{t}
\eeqn

Since $\beta$ and $\delta t$ are not affected by $\mathcal{Q}$, we can
group the last equation as 

\beq
\mathcal{Q}(-t_{i+1}){u}_{i+1} = \mathcal{Q}(-t_i)\left[{u}_i +  \beta_i {\omega}_i\delta{t}\right]
\eeq

Applying $\mathcal{Q}(t_i)$ to the former and $\mathcal{Q}(t_{i+1})$ to
the latter equation, we have 

\beqn
{\omega_i} &=& \alpha_i \mathcal{Q}(\delta t_i){\omega}_{i-1} + {f}_i\\
{u}_{i+1} &=& \mathcal{Q}(\delta t_{i+1})\left[{u}_i + \beta_i \omega_i\delta{t}\right]
\label{eq:rk32n}
\eeqn

\noindent where $\delta t_i = t_i - t_{i-1}$. 

Equation~\ref{eq:rk32n} states that in order to evolve the
system in time, one can use the RK3-2N scheme unmodified for
everything but the shear advection, and then shift both $u$ and
$\omega$ (the variable and derivative arrays) in the $\phi$ direction by
an amount $\Delta{\phi} = -\bar{u}_\phi\delta{t_i}/r$ at the end of
the $i$th stage. The procedure to translate and interpolate the variables and
derivative arrays for each stage to and from Fourier space is a costly
one, but this cost is compensated for by the increase in timestep. The increase in time step in general,
for disk advection problems, is about a factor of 10. The FFT operations
lead to a factor of 3 overhead. The overall gain is thus about a factor of 3 in performance.

The same operations are performed in the shearing sheet in
the SAFI algorithm. Notice that this proves that the SAFI
  algorithm is third-order accurate, something that was not clear from
  \cite{2009ApJ...697.1269J}. Therefore, our algorithm can be seen as the
  polar coordinate generalization of SAFI for global disks.

\subsection{Demonstration of Third-Order Accuracy in Time}

To demonstrate the third-order accuracy of our modified time integration scheme as proved in
\sect{sect:thirdorderproof} we present a simple test.
A one-dimensional problem in $\phi$ is solved
with a homogeneous fluid  accelerated in $\phi$ by a force $F_\phi(t) = \rho\, t^4$ with time $t$.
The exact solution for the fluid velocity is then $u_\phi(t) = t^5/5$. 
Running this test to $t=1$ where $u_\phi(t)=1/5$, we plot the relative error in the solution for the fluid velocity
and find third-order convergence, as shown in 
\fig{fig:timetest}.
This shows that our algorithm for splitting of the $\phi$ advection does indeed preserve 
the third-order accuracy of the {\sc Pencil Code} time integration scheme.
We remind the reader that this test isolates the time discretization, and that the spatial discretization in
 the {\sc Pencil code} is sixth-order accurate in space in most modules.

\section{Tests}
\label{sect:tests}

\begin{figure*}
 \begin{center}
   \resizebox{.8\textwidth}{!}{\includegraphics{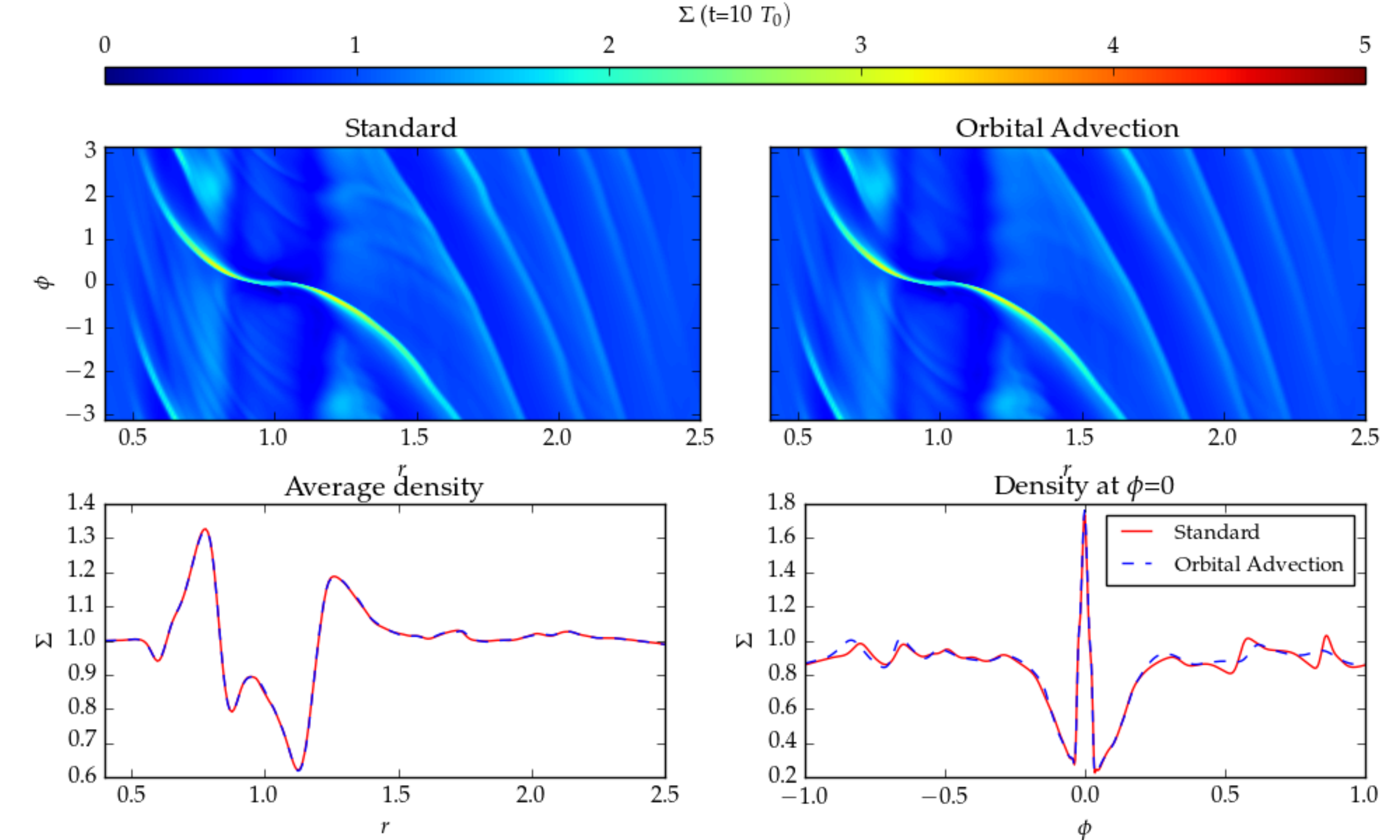}}
   \resizebox{.8\textwidth}{!}{\includegraphics{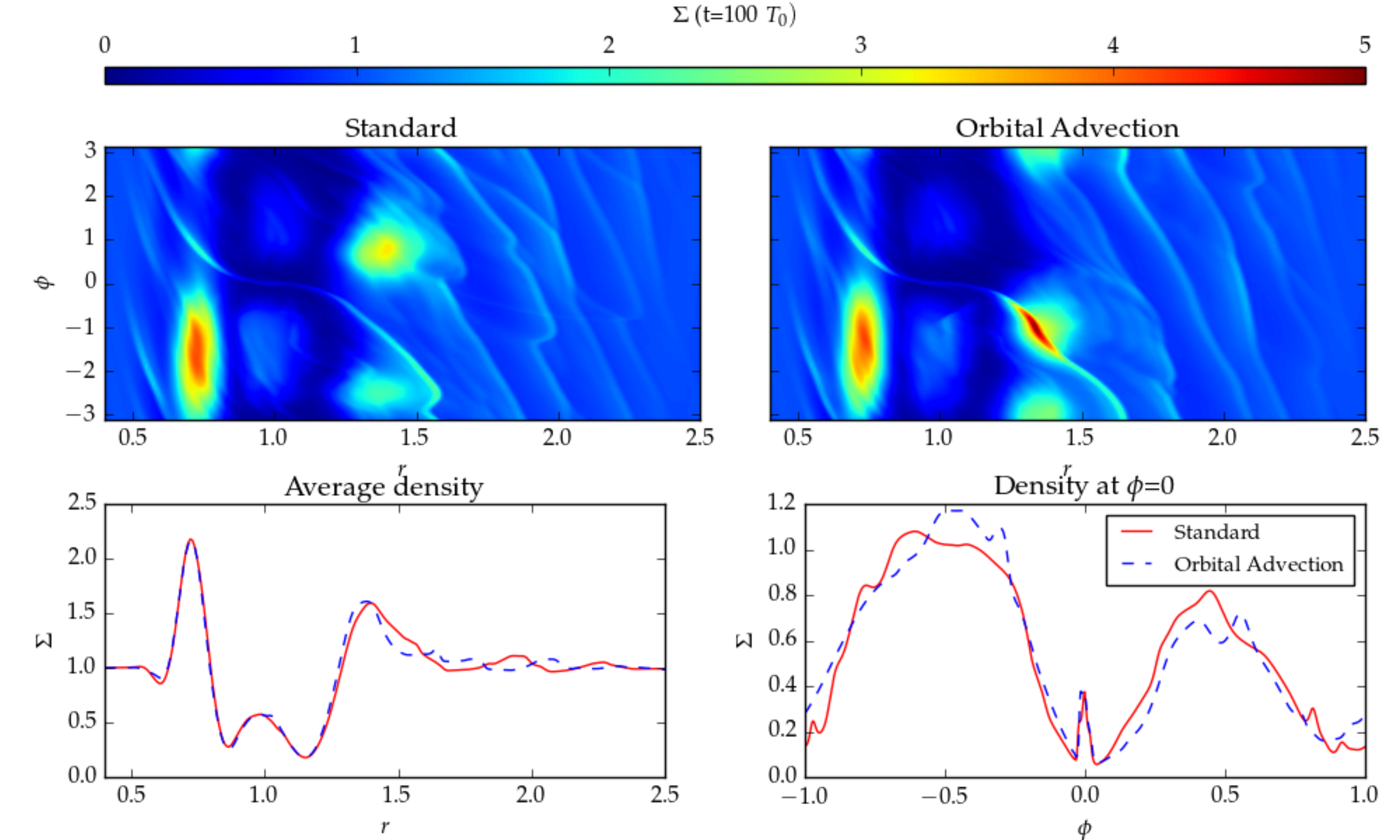}}
 \end{center}
 \caption[]{Comparison between runs with and without the orbital
   advection algorithm. Clockwise from the upper
   left: the run without the algorithm, the run with the algorithm, an
   azimuthal slice at the radial position of the planet, and the
   azimuthally averaged density as a function of radius. The upper plots show the state of the flow at
   10 orbits, the lower ones at 100 orbits. The flow at 10 orbits is
   very similar in both cases. At 100 orbits the higher numerical
   dissipation of the scheme without orbital advection has made the
   excitation of the Rossby wave instability at the outer edge
   slightly different, obviated by the different azimuthal position of
   the vortices, which hint at a different time of excitation. The
   flow is otherwise similar in both cases.}
 \label{fig:planet-fargo}
\end{figure*}

\begin{figure*}
 \begin{center}
   \resizebox{\textwidth}{!}{\includegraphics{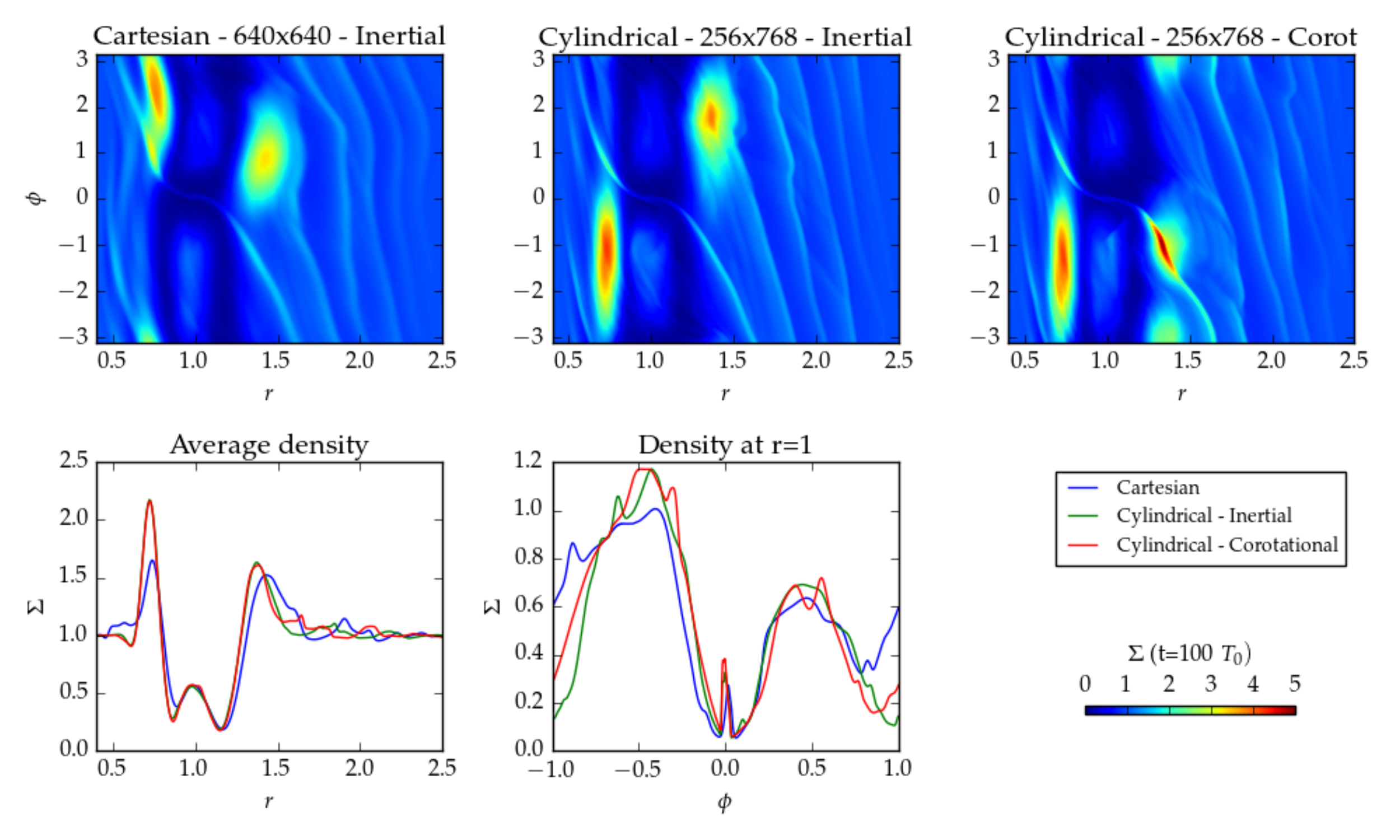}}
 \end{center}
 \caption[]{State of the flow at 100 orbits of a Jupiter-mass
   planet. The left panels show a Cartesian calculation at resolution
   640$\times$640. The cylindrical calculations are done in the
   inertial and corotational frame, which should be identical except
   for the timestep. Both use the orbital advection algorithm. The
   resolution in the Cartesian run is similar to the cylindrical
   ones. The evolution of outer vortices is different in the three
   realizations, which hints at differences in the excitation of the
   RWI due to different numerical dissipation in the different methods.}
 \label{fig:planet-fargo-inertial-vs-corot}
\end{figure*}

\begin{figure*}
 \begin{center}
   \resizebox{\textwidth}{!}{\includegraphics{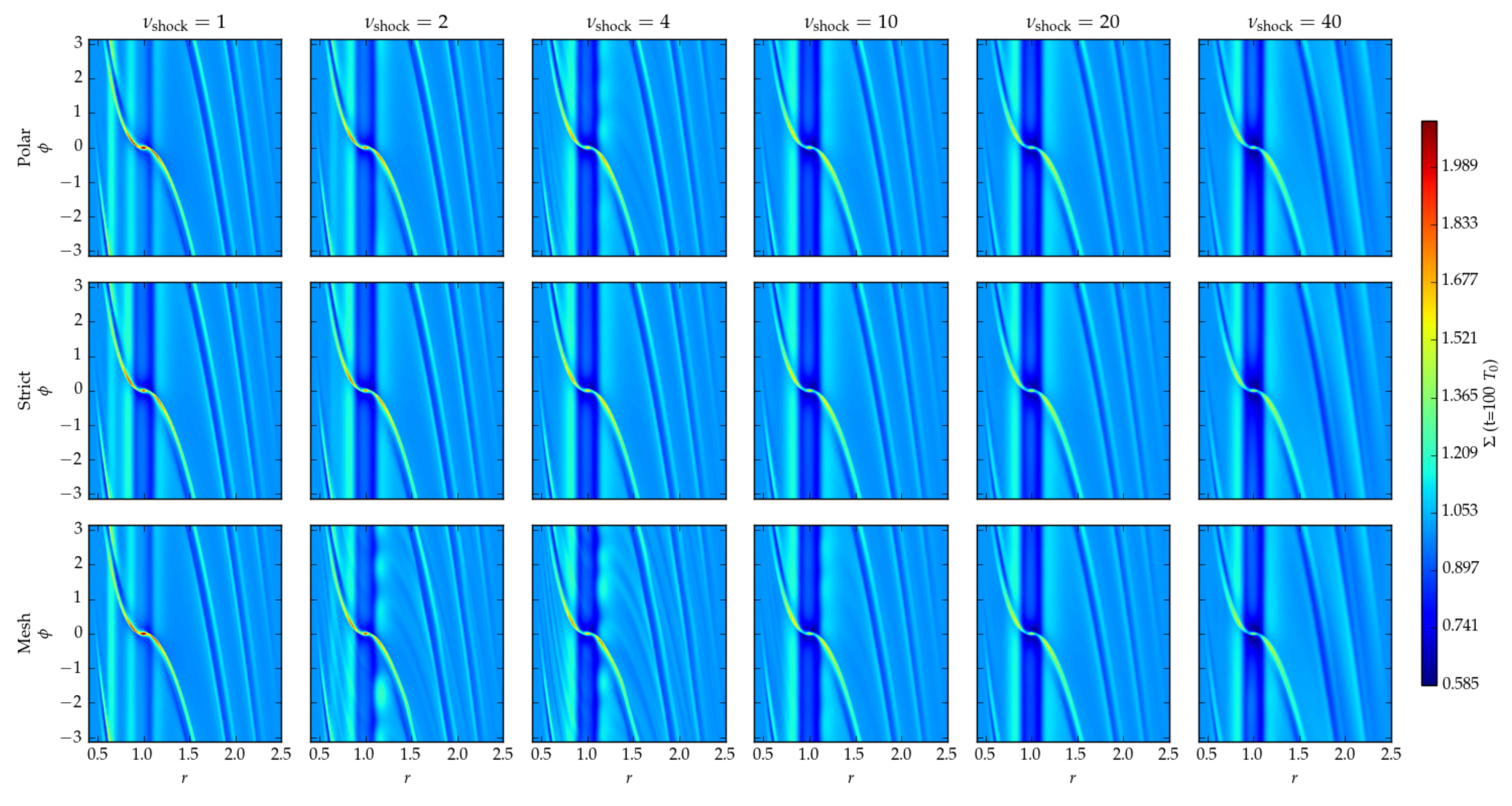}}
 \end{center}
 \caption[]{Testing the different hyperdissipation schemes. The
   rows test different shock viscosity coefficients. The columns test different schemes:  {\it polar}
   hyperdiffusion and {\it mesh} hyperdiffusion consider only the $\nabla^6$
   term. Polar scales as $\delta x^4$, whereas mesh is resolution
   independent, scaling as $\delta x^5$. {\it Strict} solves the
   $\nabla^2\nabla^2\nabla^2$ formulation. Judging from the excitation
   of the RWI at the outer gap edge, a shock viscosity of 4 is the best choice,
   working on both polar and mesh. Mesh also works well at other
   values of shock viscosity (2 and 10, respectively).}
 \label{fig:allplots_neptune}
\end{figure*}

\begin{figure*}
 \begin{center}
   \resizebox{\textwidth}{!}{\includegraphics{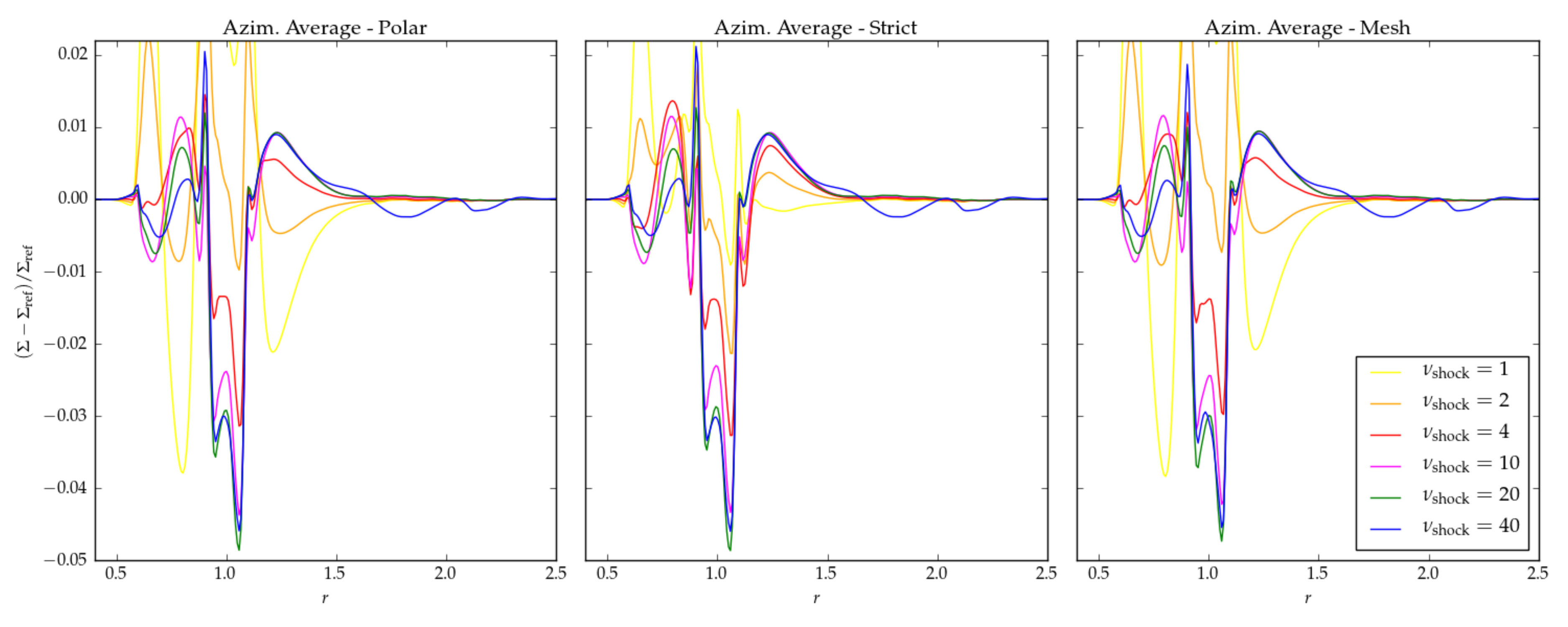}}
 \end{center}
 \caption[]{Comparison between the different hyperdiffusion methods at
   lower resolution (256$\times$768) against a reference solution at higher
   resolution (1024x3072). Shock viscosities of 1 and 2 show large
   deviations from the reference solution not only at the gap but also
   at the edges, with mass concentration in the inner edge. The gap
   gets progressively shallower as the shock viscosity increases,
   until converging at $\nu_{\rm shock}=20$. This high value of
   artificial viscosity is undesirable, and we use $\nu_{\rm shock}=4$ instead.}
 \label{fig:lineplot}
\end{figure*}

\begin{figure*}
  \begin{center}
    \resizebox{\textwidth}{!}{\includegraphics{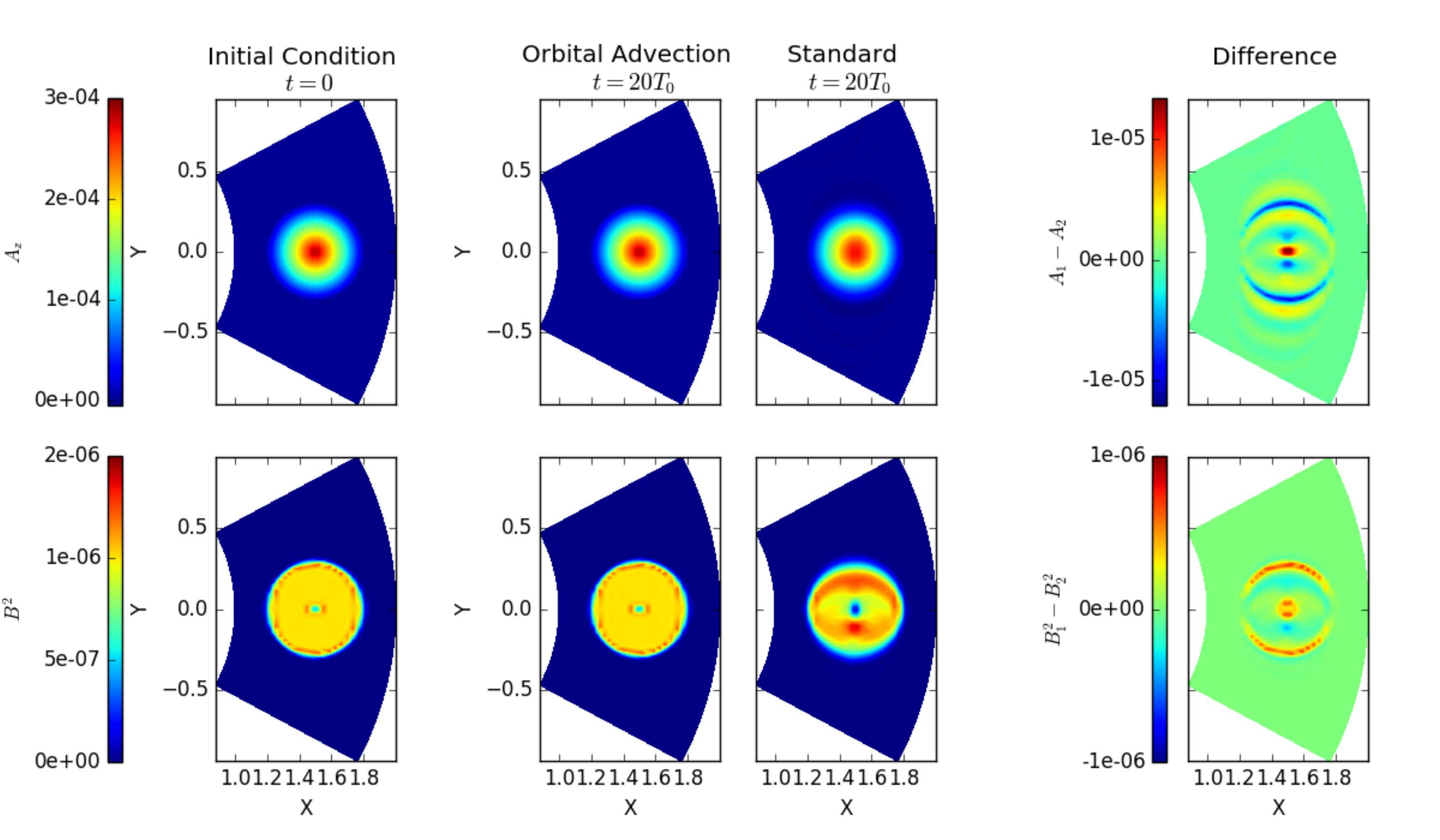}}
  \end{center}
  \caption[]{Test of the MHD orbital algorithm by advection of a field
    loop, a cylindrical coordinate adaptation of the test of
    \citet{2005JCoPh.205..509G}. 
    The upper panels show the magnetic potential, the
  lower ones the magnetic energy. The leftmost panels show the initial
  condition. The middle panels show the 
  results after 20 revolutions of the field loop. The center-left panels
  are calculated with the orbital advection algorithm, the center-right ones without it. The
  rightmost panels show the difference between the results. The calculation without
  orbital advection results in significantly more numerical diffusion:
  one revolution takes about two timesteps with the algorithm, whereas the same
  time corresponds to over 200 timesteps otherwise.}
  \label{fig:fieldloop}
\end{figure*}

\begin{figure}
\includegraphics[width=\columnwidth]{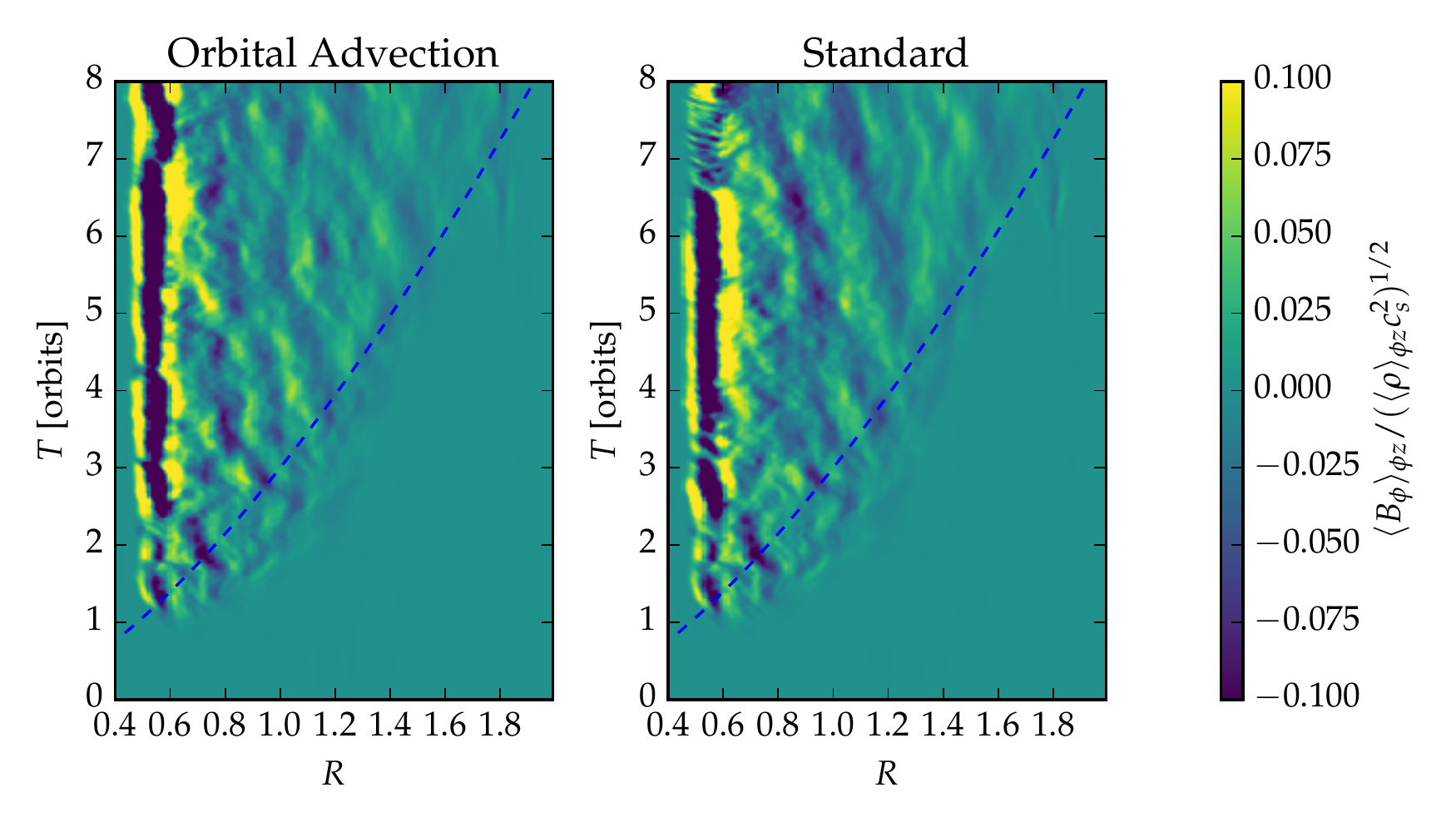}
\caption{Comparison of the initial evolution of the MRI in a cylindrical domain.
Shown are runs with our new azimuthal advection (left) and without (right). The blue dashed line denotes three local orbital periods at each radius.}
\label{fig:mricomp}
\end{figure}

\begin{figure}
\includegraphics[height=0.9\textheight]{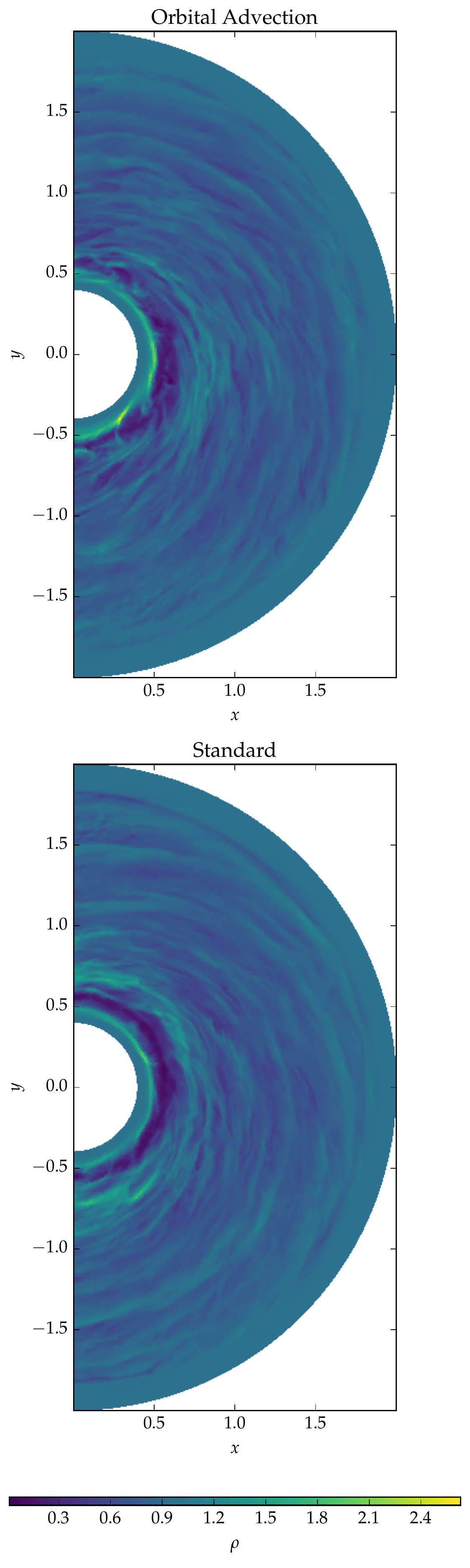}
\caption{Comparison of the evolved state of the MRI in a cylindrical domain, being a cut of the gas 
density field $\rho$ at $T=8$ orbits.
Shown are runs with our new azimuthal advection (top) and without (bottom).}
\label{fig:mrirhocut}
\end{figure}

We present in this section a series of tests aimed at describing the practical application of the scheme.
By design, the {\sc Pencil Code} relies on some high-frequency filter, in the form of a dissipation or diffusion operator,
to stabilize the scheme by eliminating structures close to the grid scale.
In our first test, a hydrodynamic planet-disk interaction problem, significant attention will be paid 
to this selection, which differs from that needed for the same problem on a Cartesian grid, 
and may vary from that needed on a cylindrical grid without the use of or orbital advection scheme.
We then proceed to show a classic magnetic field advection test, the
field loop advection problem (modified for cylindrical coordinates), 
and demonstrate the use of orbital advection in a global disk simulation of magnetorotational instability (MRI) driven turbulence.

\subsection{Planet-Disk Problem}
\label{sect:planetdisk}

We now address the planet-disk interaction benchmark problem
described in \citet{2006MNRAS.370..529D}. In that paper {\sc Pencil}
presented only the viscous runs, for Neptune and Jupiter mass (mass ratios
$q=\ttimes{-4}$ and $q=\ttimes{-3}$, respectively). Also, the runs
were performed in Cartesian coordinates; the cylindrical version of 
{\sc Pencil} was coded in a later date \citep{2009A&A...497..869L}

In a single processor at resolution
$N_r\times N_\phi$ = 128$\times$384, we perform one orbit in 90\,s, or $\xtimes{2.5}{-2}$\,hr. The 100 orbits of the benchmark
thus takes 2.5\,hr. Even considering the factor $\approx$2 in resolution (320$^2$ Cartesian
versus 128$\times$384 in cylindrical), this is a significant speed up compared to the
original 36\,hr that {\sc Pencil} took in the original code comparison
paper. The test was done in a 1.6 GHz Intel Core i5 processor. 

\Figure{fig:planet-fargo} shows a comparison between two
simulations differing only in the use of the orbital advection
algorithm. The upper plots show the state of the flow at 10 orbits,
whereas the lower plots show the state of the flow at 100 orbits. The resolution was $256 \times 768$ in cylindrical
coordinates. The simulation was done in a frame centered at the star
and corotating with the planet. Coriolis and centrifugal forces, as
well as the indirect term, were added to the simulation accordingly \citep[e.g.][]{2016ApJ...817..102L}. The velocity that enters in the Coriolis force is the full
velocity.  

It is seen that at 10 orbits the flows with and without the orbital
advection algorithm are very similar. Differences are expected as the
flow with orbital advection executes fewer timesteps and therefore has
less numerical diffusion. Indeed, at 100 orbits the flows already show
slight differences in the excitation of the Rossby vortices at the
outer gap edge. These vortices result from Rossby wave instability 
\citep[RWI,][]{1999ApJ...513..805L}
and are an expected result of disk-planet 
interaction \citep{2003ApJ...596L..91K}.

Although the overall shape of the gap is very similar
in both runs, numerical dissipation is indeed an issue that affects
the excitation of the RWI, as also shown in 
\fig{fig:planet-fargo-inertial-vs-corot}. In this figure, we
test the different methods we use to model planet-disk interaction: a
Cartesian grid (inherently without the orbital advection algorithm), a cylindrical
calculation in the inertial frame, where the bodies are evolved with a
built-in $N$-body code, and  the cylindrical corotational frame as used
in \fig{fig:planet-fargo}. The resolution element in the Cartesian run
at 640$\times$640 resolution is approximately the same as in the cylindrical
run at 256$\times$768. We see that
numerical dissipation led to markedly different evolutions of the
vortices, the shape of the gap, and the Lagrangian clouds as well. 

If the numerical dissipation we do not control modifies the results of
the simulation, then the dissipation we do control should also have an
effect. Prompted by this thought, we investigate the results of
planet-disk interaction we get under different explicit dissipation. 
Explicit dissipation terms appear in the code as
hyperviscosity and hyperdiffusion, functioning as high-frequency filters. We also employ a shock-capturing dissipation in the form of a
bulk (artificial) viscosity term. 

\subsubsection{High frequency filter}

In the {\sc Pencil Code} we use three different varieties of
high-frequency filters, that we call {\it strict}, since it strictly
obeys a conservation law, {\it polar} hyperdiffusion and {\it mesh}
hyperdiffusion. Polar and mesh hyperdiffusions are not conservative,
and differ in the way they scale with the grid element, mesh hyperviscosity
being independent of resolution. The appendix details their
implementation. 

We check the different hyperviscosity formulations in the panels of
\fig{fig:allplots_neptune}. The chosen problem is the benchmarked de Val-Borro
planet-disk interaction (see \sect{sect:planetdisk}), a classical fargo problem. 
For the polar, strict, and mesh formulations, the coefficients used were 5e-4, 3e-14, and 20, 
respectively, which should give roughly the same amount of dissipation at $r=1$ {\footnote{For reproducibility, we quote the exact code options. For the mesh formulation, the options are 
{\tt idiff='hyper3-mesh', diffrho\_hyper3\_mesh=20} and
{\tt ivisc='hyper3-mesh', nu\_hyper3=20}. 
For the strict formulation, the options are 
{\tt idiff='hyper3-strict', diffrho\_hyper3=3e-14} and  {\tt ivisc='hyper3-mu-strict-onthefly', nu\_hyper3=3e-14}.
For the polar formulation, the options are {\tt idiff='hyper3-cyl', diffrho\_hyper3=5e-4} and {\tt ivisc='hyper3-cyl',nu\_hyper3=5e-4}. The version of the code used was \#87ac0e5 on github (https://github.com/pencil-code).}}.

The figures
show that the mesh hyperdiffusion formulation is superior, with shock
viscosity $\nu_{\rm shock}$=4 best for all cases, allowing for better
excitation of the gap edge vortices. That level of shock viscosity is
needed is because of the high compressibility due to gap
carving. 

We construct now a high-resolution simulation where both numerical and artificial dissipation are
minimized, and use it as a reference to benchmark the solutions at lower
resolution that are more affected by high-frequency filters. In \fig{fig:lineplot} we show the azimuthal average of each of the
runs, plotted against this reference solution. In each plot the average
density for different shock viscosities is shown, for each
hyperviscosity method. The reference solution
is computed at resolution 1024$ \times $3072, 4 times the usual resolution, so
that the effect of hyperviscosity is diminished by a factor of 1000. 
The different runs show that increasing the shock viscosity has a
non-monotonic effect on the gap structure. Shock viscosities of 1 and
2  (yellow and orange lines) show a large departure from the gap
shape, as well as an accumulation next to the inner boundary. As the
shock viscosity is increased, the gap profile becomes increasingly shallower than
the reference value. Yet, when the shock viscosity is increased from 20 to
40, the behavior stops being monotonic, as the gap either converges at
the shock 20 level or gets slightly less shallow than it. However, such high
values of an artificial coefficient such as shock viscosity should be
avoided; it is seen in \fig{fig:allplots_neptune} that the vortices at gap edges are
quenched at the polar and mesh formulation at this high level of shock
viscosity. We thus decide on
$\nu_{\rm shock}$ = 4 as a good compromise between stabilizing the numerics and
getting the physics correct.

\subsection{Field loop advection}

We test the advection algorithm with a field loop advection similar to
that in \citet{2005JCoPh.205..509G}, adapted to cylindrical
coordinates. Formally, this is an MHD test,  and is commonly use to test constrained transport schemes 
for the magnetic induction equation when written in terms of the magnetic field.
However, it is a significantly simpler problem for the vector potential formulation of
 the induction equation, as a constrained transport scheme is not needed to ensure a divergence-free magnetic field in this case.
 Here, this test essentially demonstrates the advection errors generated by the scheme.

We perform the test in an $N_r \times N_\phi = 32\times 64$ grid. The
range in radius is from 1 to 2, and the range in azimuth from -0.5 to
0.5. The radial boundary is non-periodic and ``frozen,'' i.e., the
boundary  values are fixed at the initial condition. The velocity is initialized
in rigid rotation in the azimuthal direction, with $\varOmega=1$. An
acceleration $\v{g} = -\varOmega^2 \v{r}$ is added to maintain 
centrifugal balance. The sound speed $c_s=0.01$, and the adiabatic
index $\gamma=1$. 

The field loop is a patch of constant magnetic energy with sharp
edges. In terms of the magnetic potential, it is 

\beq
\v{A} = \mathrm{max}\left[A_0\left(0.3 - d\right),0\right] \hat{\v{z}}
\eeq

\noindent where $d= |\v{r}-\v{r}_0|$, with $|\v{r}_0|=1.5$ the location
of the center of the loop. We use $A_0=10^{-3}$.

The left panels of \fig{fig:fieldloop} show the initial condition for
the magnetic potential
(upper) and magnetic energy (lower). The middle panels show the result
after 20 periods of evolution with (middle left) and without (middle right) the orbital advection
algorithm. For the simulation with the orbital advection algorithm, at this resolution
one period takes only two timesteps. With standard advection, one period takes about 200
timesteps, which adds considerably more diffusion. The difference between the cases with and without the algorithm is
shown in the rightmost panels. The continuity, momentum, and induction equations are solved. No explicit hyperdiffusion is added.  
 

\subsection{MRI in a Cylindrical Domain}

To demonstrate the use of the induction equation, we show here a simulation that produces
MRI in a cylindrical geometry.
This  configuration is a derivative of that used in \citet{2012ApJ...756...62L}, 
solving the same MHD equations in a similar domain.
The cylindrical grid has a radial domain $r = [0.4,2.0]$,
azimuthal domain $\phi = [-\pi/2,\pi/2]$,
vertical domain $z=[-0.1,0.1]$ and resolution $N_r \times N_\phi \times N_z = 384 \times 192 \times 64$.
A net vertical field is imposed, with a  radial profile set so that three $\lambda_{\rm max}$ fit 
in the domain vertically at each radius, where $\lambda_{\rm max}$ is the most 
unstable wavelength of the linear MRI. 
The density is constant $\rho=1$ and the gas is isothermal with constant sound speed, $c_s=0.1$,
and the gas initially rotates in Keplerian equilibrium with angular velocity $\Omega=r^{-1.5}$.
The inner and outer boundary conditions are antisymmetric zero value for the 
radial velocity, zero second derivative for the azimuthal velocity, symmetrical for the vertical velocity, 
symmetrical for the density, and zero second derivative for all components 
of the magnetic vector potential.
Noise at the level $10^{-4}$ is added in the radial region $[0.6,1.8]$, and 
to stabilize the scheme, {\tt hyper3-mesh} type diffusivities on all fields set with 
the coefficient $40$ and $\nu_{\rm shock}=5$. 
We employ \citet{2006MNRAS.370..529D} buffer zones with radial width $0.1$ 
and driving timescale $0.1$ local orbital periods on both radial boundaries.
In \fig{fig:mricomp} we show the evolution of the $\phi,z$-averaged azimuthal 
magnetic field, normalized by the $\phi,z$-averaged pressure.
The two panels show that the initial growth and saturation behavior are equivalent in the two formulations.
Additionally, \fig{fig:mrirhocut} shows the gas density $\rho$ at $T=8$ orbits for both runs.

\section{Conclusions}
\label{sect:conclusions}

We have in this paper constructed an orbital advection algorithm
for global MHD simulations using the vector potential, suitable for
high-order time discretization. The algorithm relies on Fourier interpolation of the Keplerian
  advection term, as in the SAFI algorithm. The main differences between our
  algorithm and SAFI are the treatment of the induction equation, and
  that ours is generalized to polar coordinates. The magnetic potential evolution
relies on expanding the electromotive force and split the resulting
advection term into average orbital velocity and perturbation, the
former solved implicitly as with the other variables. Although an
advection term is required, this formulation does not require a gauge
transformation to an advective gauge since the remainder of
the induction equation, a shear term $u_j \partial_i A_j$, can be kept
as it is in the right-hand side. Without orbital advection, this term
would contribute terms that cancel the advection, but with orbital
advection solved implicitly, the cancellation is avoided. This circumvents the presence of
irrotational terms in the induction equation that spoil the quality
of the numerical solution in the advective gauge \citep{2011PhPl...18a2903C}. 

We prove analytically that the scheme is third-order accurate in time,
as the rest of the code, also showing that this accuracy is achieved
numerically. As a consequence, we prove that SAFI is 
  3rd order accurate as well. We test the implementation with a standard field loop
advection test \citep{2005JCoPh.205..509G} modified for the cylindrical
grid geometry, planet-disk interaction as benchmarked in \cite{2006MNRAS.370..529D}
and, finally, we test the excitation of the
MRI, one of the main applications of MHD in
disk physics. The \cite{2006MNRAS.370..529D} test is of particular
importance as it allowed us to test different high-frequency filters
and artificial viscosity schemes in order to correctly reproduce the
gap shape, as documented in the appendix. In the future we will apply
the algorithm to 3D spherical coordinates with stratification, as in \cite{2016ApJ...817..102L}. 

This scheme is self-consistent and elegant in its simplicity, allowing for a
implementation without elaborate numerical tricks. Its relatively effortless
treatment of magnetic field and accuracy makes it particularly efficient for
straightforward finite-difference methods.

\acknowledgments{
W.L. acknowledges support of Space
Telescope Science Institute through grant HST-AR-14572 and the NASA
Exoplanet Research Program through grant 16-XRP16\_2-0065. C.P.M. 
acknowledges support of an STFC Consolidated grant awarded to the QMUL Astronomy Unit 2012-2016.
This research was supported in part by the National Science Foundation
under Grant No. NSF PHY11-25915. W.L., C.P.M., and T.H. thank the hospitality 
of the Kavli Institute for Theoretical Physics in Santa Barbara, CA, where 
this work was developed. The simulations presented in this paper utilized the Stampede
cluster of the Texas Advanced Computing Center (TACC) at The
University of Texas at Austin, and the Comet cluster at the University
of California at San Diego, both through XSEDE grant TG-AST140014.
Simulations were also run at the Queen Mary's MidPlus computational 
facilities, supported by QMUL Research-IT and funded by EPSRC grant EP/K000128/1.
This work was performed in part at the Jet Propulsion Laboratory, California Institute of Technology.}

\bibliographystyle{apj}
\bibliography{fargoMHD} 

\begin{thebibliography}{33}
\expandafter\ifx\csname natexlab\endcsname\relax\def\natexlab#1{#1}\fi

\bibitem[{{Brandenburg}(2010)}]{2010MNRAS.401..347B}
{Brandenburg}, A. 2010, \mnras, 401, 347

\bibitem[{{Brandenburg} \& {Dobler}(2002)}]{2002CoPhC.147..471B}
{Brandenburg}, A., \& {Dobler}, W. 2002, Computer Physics Communications, 147,
  471

\bibitem[{{Brandenburg} \& {Dobler}(2010)}]{2010ascl.soft10060B}
---. 2010, {Pencil: Finite-difference Code for Compressible Hydrodynamic
  Flows}, Astrophysics Source Code Library

\bibitem[{{Brandenburg} {et~al.}(1995){Brandenburg}, {Nordlund}, {Stein}, \&
  {Torkelsson}}]{Brandenburg+95}
{Brandenburg}, A., {Nordlund}, A., {Stein}, R.~F., \& {Torkelsson}, U. 1995,
  \apj, 446, 741

\bibitem[{{Brandenburg} \& {Sarson}(2002)}]{2002PhRvL..88e5003B}
{Brandenburg}, A., \& {Sarson}, G. 2002, Physics Review Letters, 88, 055003

\bibitem[{{Candelaresi} {et~al.}(2011){Candelaresi}, {Hubbard}, {Brandenburg},
  \& {Mitra}}]{2011PhPl...18a2903C}
{Candelaresi}, S., {Hubbard}, A., {Brandenburg}, A., \& {Mitra}, D. 2011,
  Physics of Plasmas, 18, 012903

\bibitem[{{de Val-Borro} {et~al.}(2006){de Val-Borro}, {Edgar}, {Artymowicz},
  {Ciecielag}, {Cresswell}, {D'Angelo}, {Delgado-Donate}, {Dirksen}, {Fromang},
  {Gawryszczak}, {Klahr}, {Kley}, {Lyra}, {Masset}, {Mellema}, {Nelson},
  {Paardekooper}, {Peplinski}, {Pierens}, {Plewa}, {Rice}, {Sch{\"a}fer}, \&
  {Speith}}]{2006MNRAS.370..529D}
{de Val-Borro}, M., {Edgar}, R.~G., {Artymowicz}, P., {Ciecielag}, P.,
  {Cresswell}, P., {D'Angelo}, G., {Delgado-Donate}, E.~J., {Dirksen}, G.,
  {Fromang}, S., {Gawryszczak}, A., {Klahr}, H., {Kley}, W., {Lyra}, W.,
  {Masset}, F., {Mellema}, G., {Nelson}, R.~P., {Paardekooper}, S.-J.,
  {Peplinski}, A., {Pierens}, A., {Plewa}, T., {Rice}, K., {Sch{\"a}fer}, C.,
  \& {Speith}, R. 2006, \mnras, 370, 529

\bibitem[{{Duffell}(2016{\natexlab{a}})}]{2016ascl.soft05011D}
{Duffell}, P.~C. 2016{\natexlab{a}}, {DISCO: 3-D moving-mesh
  magnetohydrodynamics package}, Astrophysics Source Code Library

\bibitem[{{Duffell}(2016{\natexlab{b}})}]{2016ApJS..226....2D}
---. 2016{\natexlab{b}}, \apjs, 226, 2

\bibitem[{{Gardiner} \& {Stone}(2005)}]{2005JCoPh.205..509G}
{Gardiner}, T., \& {Stone}, J. 2005, Journal of Computational Physics, 205, 509

\bibitem[{{Haugen} \& {Brandenburg}(2004)}]{2004PhRvE..70b6405H}
{Haugen}, N., \& {Brandenburg}, A. 2004, Physical Review E, 70, 026405

\bibitem[{{Johansen} {et~al.}(2009){Johansen}, {Youdin}, \&
  {Klahr}}]{2009ApJ...697.1269J}
{Johansen}, A., {Youdin}, A., \& {Klahr}, H. 2009, \apj, 697, 1269

\bibitem[{{Johnson} {et~al.}(2008){Johnson}, {Guan}, \&
  {Gammie}}]{2008ApJS..177..373J}
{Johnson}, B.~M., {Guan}, X., \& {Gammie}, C.~F. 2008, \apjs, 177, 373

\bibitem[{{Koller} {et~al.}(2003){Koller}, {Li}, \&
  {Lin}}]{2003ApJ...596L..91K}
{Koller}, J., {Li}, H., \& {Lin}, D. 2003, \apj, 596, L91

\bibitem[{{Lesur} \& {Longaretti}(2005)}]{2005A&A...444...25L}
{Lesur}, G., \& {Longaretti}, P.-Y. 2005, \aap, 444, 25

\bibitem[{{Li} {et~al.}(2001){Li}, {Colgate}, {Wendroff}, \&
  {Liska}}]{2001ApJ...551..874L}
{Li}, H., {Colgate}, S.~A., {Wendroff}, B., \& {Liska}, R. 2001, \apj, 551, 874

\bibitem[{{Li} \& {Li}(2012)}]{LILI2012}
{Li}, S., \& {Li}, H. 2012, {A Fast Parallel Simulation Code for Interaction
  between Proto-Planetary Disk and Embedded Proto-Planets: Implementation for
  3D Code}, Tech. Rep. LA-UR-12-22213, Los Alamos National Lab.

\bibitem[{{Lovelace} {et~al.}(1999){Lovelace}, {Li}, {Colgate}, \&
  {Nelson}}]{1999ApJ...513..805L}
{Lovelace}, R., {Li}, H., {Colgate}, S., \& {Nelson}, A. 1999, \apj, 513, 805

\bibitem[{{Lyra}(2009)}]{LyraPhDThesis}
{Lyra}, W. 2009, PhD Thesis

\bibitem[{{Lyra} {et~al.}(2009){Lyra}, {Johansen}, {Zsom}, {Klahr}, \&
  {Piskunov}}]{2009A&A...497..869L}
{Lyra}, W., {Johansen}, A., {Zsom}, A., {Klahr}, H., \& {Piskunov}, N. 2009,
  \aap, 497, 869

\bibitem[{{Lyra} \& {Mac Low}(2012)}]{2012ApJ...756...62L}
{Lyra}, W., \& {Mac Low}, M.-M. 2012, \apj, 756, 62

\bibitem[{{Lyra} {et~al.}(2016){Lyra}, {Richert}, {Boley}, {Turner}, {Mac Low},
  {Okuzumi}, \& {Flock}}]{2016ApJ...817..102L}
{Lyra}, W., {Richert}, A., {Boley}, A., {Turner}, N., {Mac Low}, M.-M.,
  {Okuzumi}, S., \& {Flock}, M. 2016, \apj, 817, 102

\bibitem[{{Maron} {et~al.}(2012){Maron}, {McNally}, \& {Mac
  Low}}]{2012ApJS..200....6M}
{Maron}, J.~L., {McNally}, C.~P., \& {Mac Low}, M.-M. 2012, \apjs, 200, 6

\bibitem[{{Masset}(2000)}]{2000A&AS..141..165M}
{Masset}, F. 2000, \aaps, 141, 165

\bibitem[{{McNally} {et~al.}(2012){McNally}, {Maron}, \& {Mac
  Low}}]{2012ApJS..200....7M}
{McNally}, C.~P., {Maron}, J.~L., \& {Mac Low}, M.-M. 2012, \apjs, 200, 7

\bibitem[{{Mignone} {et~al.}(2012){Mignone}, {Flock}, {Stute}, {Kolb}, \&
  {Muscianisi}}]{2012A&A...545A.152M}
{Mignone}, A., {Flock}, M., {Stute}, M., {Kolb}, S.~M., \& {Muscianisi}, G.
  2012, \aap, 545, A152

\bibitem[{{Mignone} {et~al.}(2010){Mignone}, {Tzeferacos}, {Zanni},
  {Tesileanu}, {Matsakos}, \& {Bodo}}]{2010ascl.soft10045M}
{Mignone}, A., {Tzeferacos}, P., {Zanni}, C., {Tesileanu}, O., {Matsakos}, T.,
  \& {Bodo}, G. 2010, {PLUTO: A Code for Flows in Multiple Spatial Dimensions},
  Astrophysics Source Code Library

\bibitem[{{Mu{\~n}oz} {et~al.}(2014){Mu{\~n}oz}, {Kratter}, {Springel}, \&
  {Hernquist}}]{2014MNRAS.445.3475M}
{Mu{\~n}oz}, D.~J., {Kratter}, K., {Springel}, V., \& {Hernquist}, L. 2014,
  \mnras, 445, 3475

\bibitem[{{Springel}(2010)}]{2010MNRAS.401..791S}
{Springel}, V. 2010, \mnras, 401, 791

\bibitem[{{Stone} \& {Gardiner}(2010)}]{2010ApJS..189..142S}
{Stone}, J.~M., \& {Gardiner}, T.~A. 2010, \apjs, 189, 142

\bibitem[{{Umurhan} \& {Regev}(2004)}]{2004A&A...427..855U}
{Umurhan}, O.~M., \& {Regev}, O. 2004, \aap, 427, 855

\bibitem[{{Williamson}(1980)}]{Williamson80}
{Williamson}, J.~H. 1980, Journal of Computational Physics, 35, 48

\bibitem[{{Yang} \& {Krumholz}(2012)}]{YangKrumholz12}
{Yang}, C.-C., \& {Krumholz}, M. 2012, \apj, 758, 48

\end{thebibliography}

\appendix

We document here the formulation of
high-frequency filters as used in the {\sc Pencil Code} in polar
coordinates. The {\it strict} formulation is new and coded for the 
purpose of this work. The {\it polar} formulation is recapitulated from the discussion 
of hyperviscosity given in \cite{LyraPhDThesis}. The {\it mesh} formulation 
was implemented by A. Brandenburg (2017, private communication). 

\begin{figure}
\begin{center}
  \includegraphics[width=0.7\columnwidth]{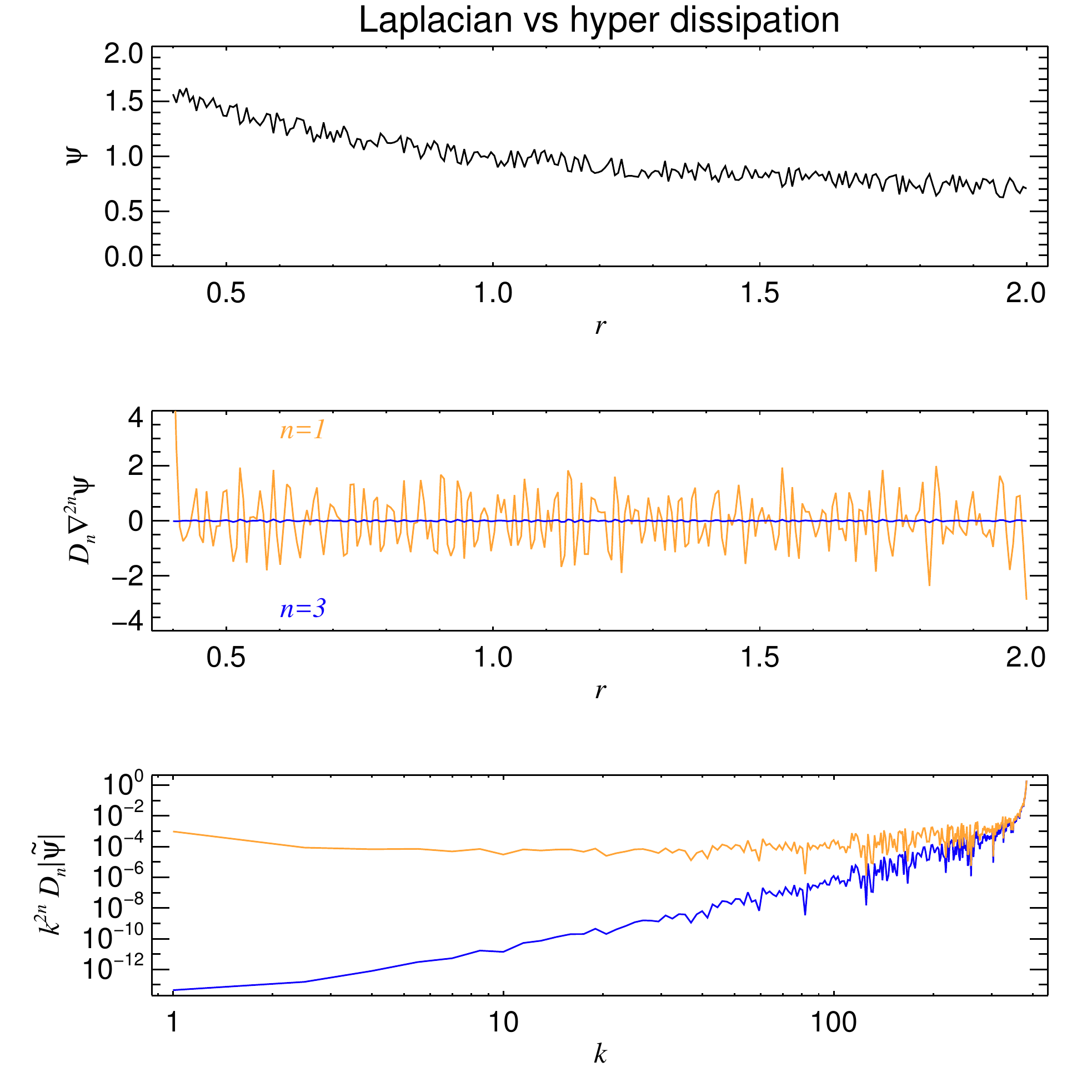}
\end{center}
\caption{Dissipation acting on a scalar field $\psi$, for $n$=1 (Laplacian dissipation) and $n$=3 (third-order hyperdissipation). The field is initially seeded with noise (upper panel). For $n$=3 the large scale is not affected as much as in the $n$=1 case, which is seen by the larger wiggling of the latter in the middle panel. In Fourier space (lower panel) we see that near the grid scale
both formulations give strong dissipation. It also illustrates that at the large scales ($k$$\simeq$1), the effect of $n$=3 is indeed negligible. The figure is reproduced from \cite{LyraPhDThesis}.}
\label{fig:hypervisc}
\end{figure}

Because of the relatively high-order discretization employed, 
{\sc Pencil} has much reduced numerical dissipation compared to an analogous low-order scheme. In
order to perform inviscid simulations, high-frequency filters can be
used to provide extra dissipation for modes approaching the Nyquist
frequency, as required to stabilize the method. The usual Laplacian viscosity $\nu\Laplace{\v{u}}$ is equivalent to
a multiplication by $k^2$ in Fourier space, where $k$ is the
wavenumber. Another tool is hyperviscosity, which replaces the $k^2$
dependency by a higher power law, $k^n$, $n > 2$. The idea behind it
is to provide large dissipation only where it is needed, at the grid
scale (high $k$), while minimizing it at the largest scales of the box
(small $k$). In principle, one can use as high $n$ as desired, but in
practice we are limited by the stencil size used in the rest of the code. A multiplication by
$k^n$ is equivalent to an operator $\nabla^n$ in real space. As {\sc Pencil}
is designed to use sixth-order finite-difference stencils for the first and second derivatives, 
three ghost cells are available in each direction;
thus the sixth-order derivative is the highest that can be conveniently
computed. The hyperdissipation we use is therefore $\nabla^6$, or $k^6$ in Fourier space. \Figure{fig:hypervisc} illustrates how this tool maximizes the inertial
range of a simulation. For further and evolving information and wisdom on the use of hyperdissipation, 
the reader is referred to the manual of the {\sc Pencil Code}.{\footnote{Some of the discussion in this section is adapted from
 entries in the manual of the {\sc Pencil Code}, which were added by the authors.}}

\subsection{Strictly Conservative Hyperdissipation}

Hyperdiffusivity is meant purely as a numerical tool to dissipate
energy at small scales and comes with no guarantee that results are
convergent with regular second-order dissipation (see
\citealt{2004PhRvE..70b6405H} for a discussion). In fact, large-scale dynamo action is known to be seriously altered in simulations of closed systems where magnetic helicity is conserved: this results in prolonged saturation times and enhanced saturation amplitudes \citep{2002PhRvL..88e5003B}.

It is desirable to have the high-frequency filters obeying the
conservation laws. So, for density we want a mass-conserving term, for velocities we want a
momentum-conserving term, for magnetic fields we want a term conserving
magnetic flux, and for entropy we want an energy-conserving term.
These enter as hyperdiffusion, hyperviscosity, hyper-resistivity, and
hyper heat conductivity terms in the evolution equations.
To ensure conservation under transport, they must take the form of the
divergence of the flux {$\v{\mathcal J}$} of the quantity $\psi$, so that
the Gauss theorem applies and we have

\begin{equation}
  \pderiv{\psi}{t} + \Div{\v{\mathcal J}} = 0.
\end{equation}

For density, the flow due to mass diffusion is usually taken as the
phenomenological Fick's law

\begin{equation}
  \label{eq:fick}
  \v{\mathcal J} = -D \grad \rho
\end{equation}

\noindent i.e., proportional to the density gradient, in the opposite direction.
This leads to the usual Laplacian diffusion

\begin{equation}
  \label{eq:diff}
  \pderiv{\rho}{t} = D\Laplace{\rho}
\end{equation}

\noindent under the assumption that the diffusion coefficient $D$ is isotropic.
Higher-order hyperdiffusion of order $2n$ involves a generalization of
\eq{eq:fick}, to

\begin{equation}
  \label{eq:hyperfick}
  \v{\mathcal J}^{(n)} = (-1)^n D^{(n)} \grad^{2n-1} \rho \; .
\end{equation}

In our case, we are interested in the case $n=3$, so that the
hyperdiffusion term is

\begin{equation}
  \label{eq:hyperdiff}
  \pderiv{\rho}{t} = D^{(3)}\nabla^6{\rho},
\end{equation}

where the operator 

\beq
\nabla^6 \equiv \nabla^2\left(\nabla^2\left(\nabla^2\right)\right).
\eeq

In Cartesian coordinates this is expanded into

\beq
\nabla^6  = \pderivn{}{x}{6} + \pderivn{}{y}{6} + \pderivn{}{z}{6} + 3\left(\pderivn{}{x}{4}\pderivn{}{y}{2} +
              \pderivn{}{x}{4}\pderivn{}{z}{2} + 
              \pderivn{}{x}{2}\pderivn{}{y}{4}  + \pderivn{}{x}{2}\pderivn{}{z}{4} +
              \pderivn{}{y}{4}\pderivn{}{z}{2} + 
              \pderivn{}{z}{4}\pderivn{}{y}{2} \right) + 
              6 \pderivn{}{x}{2} \pderivn{}{y}{2} \pderivn{}{z}{2}. 
\label{eq:del6strict}
\eeq

In cylindrical coordinates curvature terms would appear, involving
powers of $1/r$. However, if we keep only the highest-order terms,
those of order $\mathcal{O}(\Delta x^{-6}$), the curvature terms can be
discarded. The only remaining terms are those in \eq{eq:del6strict},
provided we do the trivial substitution $\partial
x\rightarrow \partial r$ and $\partial y
\rightarrow r\partial\phi$. We call this {\it strict}
hyperdiffusion, since we strictly maintain all the highest-order terms. 

\subsubsection{Hyperviscosity}

Viscosity has some caveats where subtleties apply. The difference is that the momentum flux due to viscosity is not proportional to the velocity gradient, but to the rate-of-strain tensor

\begin{equation}
  \label{eq:strain}
  S_{ij} = \frac{1}{2}
           \left(
             \pderiv{u_i}{x_j} + \pderiv{u_j}{x_i}
             - \frac{1}{3}\delta_{ij} \Div{\v{u}}
           \right) \; ,
\end{equation}

\noindent which only allows the viscous acceleration to be reduced to the simple
formulation $\nu\Laplace{\v{u}}$ under the condition of incompressibility
and constant dynamical viscosity $\mu=\nu\rho$.
Due to this, the general expression for conservative hyperviscosity
involves more terms. In the general case, the viscous acceleration is

\begin{equation}
  f_{\rm visc} = \rho^{-1} \Div{\left(2\rho\nu\Strain\right)}.
\end{equation}

So, for the hyperviscous force, we must replace the rate-of-strain tensor
by a high-order version

\begin{equation}
  \label{eq:hypervisc}
  f_{\rm visc}^{({\rm hyper})} = \rho^{-1} \Div{\left(2\rho\nu_n\Strain^{(n)}\right)}
\end{equation}

where the $n^{\rm th}$-order rate-of-strain tensor is

\begin{equation}
  \Strain^{(n)} = (-\nabla^2)^{n-1}\Strain.
\end{equation}

For the $n=3$ case it is

\begin{equation}
  S^{(3)}_{ij} = \frac{1}{2}
                 \left(
                   \pderivn{u_j}{x_i}{5}
                   + \pderivn{}{x_i}{4}
                     \left( \pderiv{u_i}{x_j} \right)
                   - \frac{1}{3} \pderivn{}{x_i}{4}
                     \left(\Div{\v{u}}\right)
                 \right) \; .
\end{equation}

Plugging this into Eq.~(\ref{eq:hypervisc}), and assuming
$\mu_3=\rho\nu_3={\rm const}$

\begin{equation}
  \label{eq:hyper-mu-const}
  f_{\rm visc}^{({\rm hyper})}
    = \nu_3 \left(
              \nabla^6\v{u}
              + \frac{1}{3}\nabla^4({\grad({\Div{\v{u}}})})
            \right) \; .
\end{equation}

For $\nu_3={\rm const}$, we have to take derivatives of density as well

\begin{equation}
  \label{eq:hyper-nu-const}
  f_{\rm visc}^{({\rm hyper})} = \nu_3 \left(
    \nabla^6\v{u} + \frac{1}{3}\nabla^4({\grad({\Div{\v{u}}})}) +
    2\Strain^{(3)}\cdot\grad{\ln\rho}.
 \right)
\end{equation}

Here we can again ignore the curvature terms as they will produce
terms of lower order than $\mathcal{O}(\Delta x^{-6})$. Therefore, the
$\nabla^6$ operator applied to a vector is equal to the sum of
$\nabla^6$ on its components, as would $\nabla^2$ in 
in Cartesian coordinates. As for the other term, we can write for the
$x$-component 

\beq
\nabla^4({\grad({\Div{\v{u}}})})_x = f(u_x) + g(u_y) + h(u_z)
\eeq

\noindent and retaining only the $\partial^6$ terms, we have 

\beq
f(u_x) = \left[ \pderivn{}{x}{6}  + 2\left(
    \pderivn{}{x}{4}\pderivn{}{y}{2} +
    \pderivn{}{x}{4}\pderivn{}{z}{2} +
    \pderivn{}{y}{4}\pderivn{}{x}{2} +
    \pderivn{}{z}{4}\pderivn{}{x}{2}\right) +4\pderivn{}{x}{2}\pderivn{}{y}{2}\pderivn{}{z}{2}\right] u_x,
\eeq

\beq
         g(u_y) =  \left( \pderivn{}{x}{5}\pderiv{}{y} +
         \pderiv{}{x}\pderivn{}{y}{5} +
         3\pderivn{}{x}{3}\pderivn{}{y}{3} + 2\pderivn{}{x}{3}\pderiv{}{y}\pderivn{}{z}{2} 
         + 3\pderiv{}{x}\pderivn{}{y}{3}\pderivn{}{z}{2}
         +\pderiv{}{x}\pderiv{}{y}\pderivn{}{z}{4}\right) u_y,
\eeq

\noindent and 

\beq
         h(u_z) =  \left( \pderivn{}{x}{5}\pderiv{}{z} +
         \pderiv{}{x}\pderivn{}{z}{5} +
         3\pderivn{}{x}{3}\pderivn{}{z}{3} + 2\pderivn{}{x}{3}\pderiv{}{z}\pderivn{}{y}{2} 
       + 3\pderiv{}{x}\pderivn{}{z}{3}\pderivn{}{y}{2}
         +\pderiv{}{x}\pderiv{}{z}\pderivn{}{z}{y}\right) u_z.
\eeq

Per symmetry, the formulation for the $y$ and $z$ components are
identical under the permutation [$xyz$]. As a result of the present work,
this formulation of hyperviscosity is implemented in the {\sc Pencil
Code}. It is exact for Cartesian coordinates and accurate to first
order in polar coordinates. 

\subsection{Non-conservative Hyperdissipation}

Equations (\ref{eq:hyper-mu-const}) and (\ref{eq:hyper-nu-const})
explicitly conserve linear {\em and} angular momentum.
Although desirable properties, such expressions are cumbersome 
due to the mixed derivatives. Yet, the spectral range in which 
hyperviscosity operates is very limited and, as a numerical tool, only 
its performance as a high-frequency filter is needed. Since the terms
are artificial, it is not clear if enforcing conservation on them is
of great utility. We try thus other hyperdiffusion formulations that
abandon strict conservation. We apply a simple hyperviscosity 

\begin{equation}
 f_{\rm visc}
   = \left\{
       \begin{array}{l r}
         \nu_3 \nabla^6\v{u}   &\mbox{if $\mu={\rm const}$} \\
         \nu_3 \left(\nabla^6\v{u}
           + 2\Strain^{(3)}\cdot\grad{\ln\rho} \right)
                             &\mbox{if $\nu={\rm const}$}
       \end{array}
     \right.
\end{equation}

Notice that this can indeed be expressed as the divergence of a simple
rate-of-strain tensor

\begin{equation}
 \label{eq:weird-strain}
 S_{ij}^{(3)} = \pderivn{u_i}{x_j}{5} \; ,
\end{equation}

\noindent so it does conserve linear momentum.
It does {\it not}, however, conserve {\it angular} momentum, since the
symmetry of the rate-of-strain tensor was dropped.
Thus, vorticity sinks and sources may be spuriously generated at the grid
scale.

A symmetric tensor can be computed that conserves angular momentum and
can be easily implemented

\begin{equation}
  \label{eq:symmetric-strain}
  S_{ij} = \frac{1}{2}
           \left(
             \pderivn{u_i}{x_j}{5} + \pderivn{u_j}{x_i}{5}
           \right).
\end{equation}


This tensor, however, is not traceless, and therefore accurate only
for weak compressibility.
It should work well if the turbulence is subsonic. We performed
simulations of the MRI with both formulations, not
finding significant differences for the turbulent problem studied. 
This supports the use of the highest-order terms only, since these
are the ones that provide quenching at high $k$. We are thus drawn to
a formulation that keeps only the $\partial^6/\partial q^6$ terms. 

\subsubsection{Polar Hyperdiffusion}

The hyperdiffusion coefficient $D^{(3)}$ in \eq{eq:hyperdiff} can be calculated
from $D$ assuming that at the Nyquist frequency the two
formulations (\ref{eq:diff}) and (\ref{eq:hyperdiff}) yield the same
quenching. Considering a wave as a Fourier series of a generalized
coordinate  ($q$),
one element of the series is expressed as

\begin{equation}
  \psi_k = A {\mathrm e}^{i(k q -  \omega t)}.
\end{equation}

Plugging it into the second-order diffusion equation (\ref{eq:diff})
we have the dispersion condition $i\omega = D k^2$.
The sixth-order version (\ref{eq:hyperdiff}) yields $i\omega = D^{(3)}
k^6$.

Equating both we have  $D^{(3)} = D k^{-4}$. This condition should 
hold at the grid scale, where $k=\pi/\Delta q$, therefore

\begin{equation}
  \label{eq:diffrho-hyper}
  D^{(3)} = D \left(\frac{{\Delta}q}{\pi}\right)^4.
\end{equation}

And the discretized equation of motion is therefore 

\beq
 \frac{\Delta\psi}{\Delta t} =D^{(3)}
 \frac{\Delta^6\psi}{\Delta q^6} = \frac{D}{\pi^4}
 \frac{\Delta^6\psi}{\Delta q^2} .
\eeq

This formulation, being dependent on line element $\Delta q$ is appropriate for
use with polar coordinates. As such, we call it {\it polar}
  hyperdiffusion. The formulation for viscosity, resistivity, and heat
conduction follows along the same lines.

\subsubsection{Resolution-independent Mesh Hyper-Reynolds Number}

The polar hyperdiffusion above scales with resolution in a way that it can
be written in terms of a Laplacian second-order coefficient. In this
way, the relationship between the second-order coefficient and the
Reynolds number scales linearly with resolution. 

One can rewrite this in terms of a coefficient that instead keeps the
hyper-Reynolds number constant with resolution. The hyper-Reynolds
number is 

\begin{equation}
\mbox{Re}_{\rm grid}=
u_{\rm rms}\left/\left(\nu_n k_{\rm Ny}^{2n-1}\right)\right.
\end{equation}

\noindent where, $k_{\rm Ny}=\pi/\Delta q$ is the Nyquist
wavenumber. The $n=3$ hyperdiffusion coefficient thus should thus be
$\nu^{(3)}$=$C \Delta q^5$, with $C = \urms/( \Rey \ \pi^5)$
as proportionality coefficient. So, the discretized equation, in a way
that strictly keeps the Reynolds number constant at the grid scale, is 

\begin{equation}
 \frac{\Delta\psi}{\Delta t} =D^{(3)}  \frac{\Delta^6\psi}{\Delta q^6}
 = \frac{\urms}{\Rey \ \pi^5} \frac{\Delta^6\psi}{\Delta q}.
\end{equation}

Because $\urms$ is a dynamical quantity, a strict retention of constant
$\Rey$ would need to have a dynamical $D^{(3)}$ as well. Instead, we use
a representative $\urms$ at start time, and write 

\begin{equation}
D^{(3)} =   \frac{D_{\rm mesh} \Delta q}{60 \pi^5} 
 = \frac{\urms}{\Rey \ \pi^5} \frac{\Delta^6\psi}{\Delta q} 
\end{equation}

\noindent so, that way, $D_{\rm mesh} /60 = \urms/\Rey$ sets the hyperdiffusion.

Like the polar hyperviscosity, this formulation sets the
hyperviscosity in a coordinate-independent way. Moreover, it has the
advantage that the Reynolds number is kept constant over the mesh as
the cells in the polar or non-uniform grid differ in size. We call
this formulation {\it mesh} hyperdiffusion for this reason. A similar choice
is implemented in the code for shearing boxes, using the maximum
velocity instead of $\urms$ and updated dynamically, described in \cite{YangKrumholz12}.

\end{document}